%% file: hydra-sol.tex
\documentclass{jfm}

\usepackage{graphicx,color}
\usepackage{latexsym}
\usepackage{amssymb}
\usepackage{amsfonts}
\usepackage{amsmath}
\usepackage{epsfig}
\usepackage{epstopdf}
\usepackage{natbib}
\title[Soliton generation by internal tidal beams impinging on a pycnocline]
{Soliton generation by internal tidal beams impinging on a pycnocline : laboratory experiments}

\author[Hydralab team]
{
Matthieu Mercier, Manikandan Mathur, Louis Gostiaux, Theo Gerkema, Jorge Magalh\~aes, Jos\'e da Silva, Thierry Dauxois.`
}

\author[M. J. Mercier and others]%
{M\ls A\ls T\ls T\ls H\ls I\ls E\ls U\ns J.\ns M\ls E\ls R\ls C\ls I\ls E\ls R$^{1,6}$%
  \thanks{Email address for correspondence: mmercier@mit.edu},\ns
M\ls A\ls N\ls I\ls K\ls A\ls N\ls D\ls A\ls N\ns M\ls A\ls T\ls H\ls U\ls R$^{2,3,1}$, \ns
L\ls O\ls U\ls I\ls S\ns G\ls O\ls S\ls T\ls I\ls A\ls U\ls X$^2$, \ns
T\ls H\ls E\ls O\ns G\ls E\ls R\ls K\ls E\ls M\ls A$^4$, \ns
J\ls O\ls R\ls G\ls E\ns M.\ns M\ls A\ls G\ls A\ls L\ls H\ls \~A\ls E\ls S$^5$, \ns
J\ls O\ls S\ls \'E\ns C.\ns B.\ns D\ls A\ns S\ls I\ls L\ls V\ls A$^{5}$\ns \and 
T\ls H\ls I\ls E\ls R\ls R\ls Y\ns D\ls A\ls U\ls X\ls O\ls I\ls S$^6$} 

\affiliation{$^1$Department of Mechanical Engineering, Massachusetts Institute of Technology,
77 Massachusetts Avenue, Cambridge, MA 02139, USA.\\[\affilskip]
$^2$ Laboratoire des Ecoulements Geophysiques et Industriels (LEGI), UMR 5519 CNRS-UJF-INPG, 21 rue des Martyrs, 38000 Grenoble, France. \\[\affilskip]
$^3$ Laboratoire de M\'et\'eorologie Dynamique, \'Ecole Polytechnique, 91128 Palaiseau, France.  \\[\affilskip]
$^4$ Royal Netherlands Institute for Sea Research (NIOZ), P.O. Box 59, 1790 AB Den Burg, Texel, The Netherlands. \\[\affilskip]
$^5$ CIMAR/CIIMAR, Centro Interdisciplinar de Investigac\~ao Marinha e Ambiental and Departamento de Geoci\^{e}ncias, Ambiente e Ordenamento do Territ\'{o}rio, Universidade do Porto, Rua do Campo Alegre, 687, 4169-007 Porto, Portugal. \\[\affilskip] 
$^6$ Laboratoire de Physique de l'\'{E}cole Normale Sup\'{e}rieure de Lyon, Universit\'{e} de Lyon, CNRS, 46 All\'{e}e d'Italie, F-69364 Lyon cedex 07, France.}

\date{\today}

\begin{document}

\maketitle

\begin{abstract}

In this paper, we present the first laboratory experiments
that show the generation of internal solitary waves by the impingement of a quasi two-dimensional internal wave beam on a pycnocline.
{\color {black} These experiments were inspired by} 
observations of internal solitary waves in the deep ocean 
{\color {black} from} Synthetic Aperture Radar (SAR) imagery, {\color {black} where this
so-called mechanism of `local generation' was argued to be at work: here in the form of internal tidal beams hitting the thermocline.}
Nonlinear processes involved here are {\color {black} found to be} of two kinds. First, we observe the generation of a mean flow and higher harmonics {\color {black} at the location where the principal beam reflects from the surface and pycnocline;}
their characteristics are 
{\color {black} examined}
 using Particle Image Velocimetry measurements. Second, the appearance of internal solitary waves {\color {black} in the pycnocline, detected with ultrasonic probes; they
are further characterized by a bulge in the frequency spectrum, distinct from the higher harmonics.}
Finally, the relevance of our results for understanding ocean observations is discussed.

%

\end{abstract}

\include{1_intro}

\include{2_description}

\include{3_flow}

\include{4_pycnocline}

\include{5_discussion}

\acknowledgements
We thank Samuel Viboud and Henri Didelle for their help during the experiments, Marc Moulin for the cams design and acknowledge helpful discussions with Joel Sommeria. {\color {black} We also thank the anonymous referees for useful suggestions concerning the mean flow}. The experiments were supported by funds from the Hydralab III Transnational Access Program (6th FP) and ANR PIWO (contract number ANR-08-BLAN-0113). We also thank MIT-France for partially funding the travel expenses of Manikandan Mathur.

\appendix
\include{A_appendix}
\include{B_appendix}

\bibliography{REF}

\end{document}

%% file: 1_intro.tex
\section{Introduction}

Internal solitary waves (ISWs) are among the most noticeable kinds of
internal waves in the ocean. Their surface manifestation may be visible
even from spacecraft~\citep[see, for example, ][]{jack07}.  They often appear in groups,
and the groups themselves usually appear regularly, every tidal
period.  This points to a tidal origin: barotropic tidal flow over
topography creates internal tides, which, while propagating away from
their source, may steepen and split up into ISWs. This is the common
picture, in which the internal tide is regarded as a horizontally
propagating, interfacial wave. In the early nineties, \cite{new90,new92}
showed the first evidence of a
very different generation mechanism of ISWs. The origin is still
tidal, but the mechanism involves an internal tidal beam,
generated over the continental slope, which first propagates downward,
then reflects from the ocean floor, and finally, as its energy goes upward,
impinges on the seasonal thermocline. Here it creates a depression, which,
while propagating away, steepens and evolves into ISWs. This interpretation has been
corroborated by Synthetic Aperture Radar (SAR) imagery, showing a great increase of ISWs in the
central Bay of Biscay, just beyond the area where the beam is expected to approach
the surface~\citep{new02,dasi07}. Recently, the same mechanism has been proposed to explain ISWs off Portugal~\citep{dasi07} and in the Mozambique
Channel \citep{dasi09}.

In this paper we present results on laboratory experiments -- carried
out at the LEGI Coriolis Platform in Grenoble in 2008 -- that were set up
to create ISWs by this very mechanism, i.e.\ by an internal wave beam
impinging on a pycnocline. At the beginning of the generation process
lies the internal `tidal' beam, generated here with a recently
designed wavemaker \citep{gost07a,merc10}, which creates a well-defined
monochromatic unidirectional beam.  We varied the stratification,
i.e.\ the strength, depth and thickness of the pycnocline which was set up above a constantly stratified lower layer. For each stratification we carried out experiments with different forcing frequencies, thus changing the angle the beam makes with the vertical. We note here that the wave beam generated by the wavemaker directly models the upward-propagating tidal beam that impinges on the ocean pycnocline.

In the set-up of the experiments, we were guided by theoretical
studies on the subject. The process starts at the wavemaker, from which the
internal-wave beam originates. So long as it propagates through the
layer of uniform stratification, no significant changes are expected
to occur, because a unidirectional wave beam propagating in a uniformly stratified Boussinesq fluid
satisfies not only the linear but also the nonlinear equations of motion \citep{taba03}. However,
when wave beams propagating in multiple directions intersect, nonlinear effects such as 
generation of higher harmonics occur \citep{taba05,jian09}. This situation occurs
when beams cross or reflect from boundaries.

In our set-up, and in the ocean too, the first deformation of the
beam is expected to happen when it leaves the layer of near-constant
stratification, i.e.\ when it encounters the pycnocline. This strong
inhomogeneity of the medium causes the beam to reflect partially from
and within the pycnocline, a linear process that `scatters' the
beam. As a result, some of the energy stays behind in the pycnocline. It has been shown
in theoretical studies that this forms the basis from which ISWs may
later evolve \citep{gerk01,akyl07}. At the same time, the partial
reflections (that occur when a unidirectional wave beam propagates through a region of
non-constant stratification) 
within the pycnocline, as well as the full reflection from
the upper surface, create junctions of crossing beams, and hence form
a source of nonlinear generation of higher harmonics. In our
laboratory experiments, we thus focused on these two features: first,
the evolution of the pycnocline after the impact of the beam, and
second, higher harmonics. In a way, they are contrasting phenomena:
ISWs in the pycnocline are the nonlinear result of a depression that
steepens, the depression itself originating from an essentially
linear process (internal reflections in the pycnocline). On the other
hand, higher harmonics waves behave as linear
waves but find their origin in a nonlinear process at the junction of
the main beam and its reflected counterpart. This distinction in
behaviour has consequences for the tools by which they may be
analysed; in particular, the application of harmonic analysis is
well-suited to extract the essentially linear higher harmonic beams
from the measurements, but cannot expect to yield anything clear on
the genuinely nonlinear ISWs. The same is true for wave spectra. We
discuss this further in sections~\ref{sec:3}~\&~\ref{sec:4}.


The linear regime of the problem we address, an internal wave beam
impinging a continuously stratified pycnocline in a finite-depth tank,
was studied in \cite{math09}. Their experimental results on wave beam
ducting, a scenario where the incident energy tends to remain trapped
in the pycnocline even after multiple reflections within the
pycnocline, were in excellent agreement with the viscous theory. This
linear process, as discussed in \cite{gerk01} and \cite{akyl07},
serves as a precursor to the formation of solitons.

Previous experimental and theoretical studies have identified the key
parameters governing the response of the pycnocline to the incident
beam. 
\cite{deli75} derived a parameter $\beta$ which they interpreted as the square of the ratio of the phase speed for interfacial waves and the horizontal phase speed of the incident beam. This interpretation is open for some debate (see \cite{thor98}). The experimental results of \cite{deli75}
show that the response of the pycnocline is largest when
$|\beta|\approx1$.  In a different setting, in which the interfacial
waves were assumed to be long rather than short, a similar criterion
was identified in terms of a parameter $\gamma$, defined as the ratio
of the phase speed of interfacial waves and that of the first mode of
the uniformly stratified lower layer \citep{gerk01}. We amplify on
this below. Finally, \cite{akyl07}, in a setting closer to that of
\cite{deli75} -- their lower uniformly stratified layer being
infinitely deep -- recovered a parameter similar to $\beta$, which
they call $\alpha$.  In either setting, the parameter $g'$ plays a
key role, and we study its influence by examining different shapes and
strengths of the pycnocline.

{\color {black} Recently, in a configuration similar to the present setup, \cite{gris11}
modelled the local generation of ISWs with a nonlinear, non-hydrostatic numerical code
(MIT-gcm) and showed that higher mode ISWs (namely mode-2 and 3) may also be
generated by this mechanism.
Initially, the idea was to carry out numerical runs concurrently with the laboratory experiments, 
and the basic setting for the numerical model was chosen to be the same as in the laboratory. 
But, as shown in the rest of this paper, we observed some important qualitative differences 
(between the experiments and the simulations) that rendered a direct comparison impossible.}

The present paper is organised as follows. In section 2, a description of the
experimental set-up and the measurement techniques is given. An
analysis of the results from Particle Image Velocimetry is given in
Section 3, where we focus on the mean flow and harmonics. Results from
the ultrasonic probes are presented in Section 4; they provide a quantitative view
of the pycnocline displacements and the ISWs. Finally, we end
with a discussion of our results and conclusions in section 5.

%% file: 2_description.tex
\section{Description of the experimental set-up}\label{setup}

\begin{figure}
\begin{center}
\includegraphics[width=0.925\linewidth]{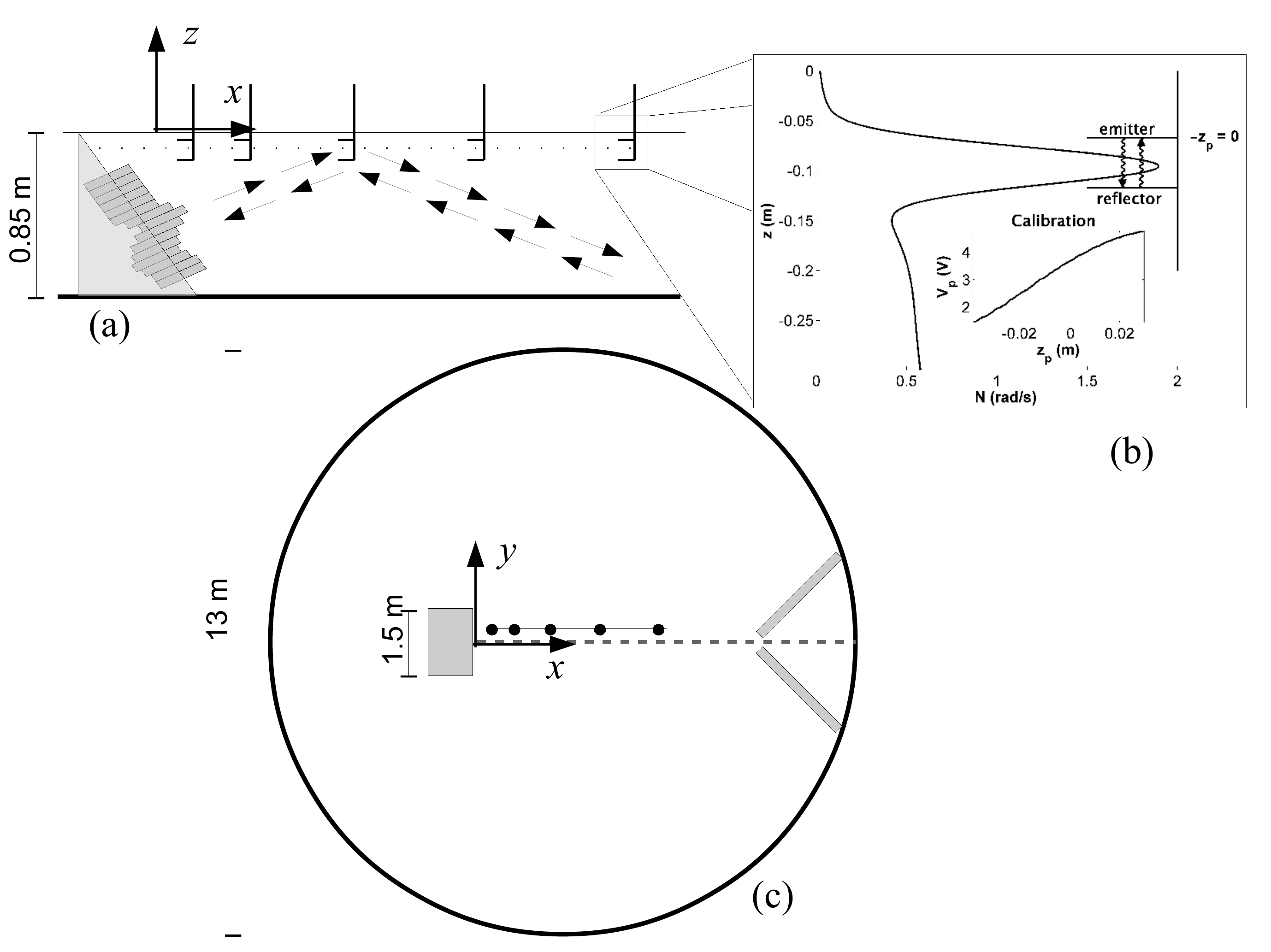}
\caption{\textcolor {black}{Schematic of the experiment, (a) side view, (b) zoom-in on the ultrasonic probe and its relative positioning with respect to the stratification, and (c) top
view. A typical calibration curve obtained by traversing the probe in quiescent medium is shown in the inset in (b).}}
\label{scheme_exp}
\end{center}
\end{figure}

The experiments were carried out at the Coriolis Platform
of LEGI, Grenoble (see Figure~\ref{scheme_exp} for a schematic of the experimental setup). This facility is a 13~m
diameter rotating basin that can be filled with salty water; during the process
of filling, the salinity can be changed, so that a vertically
stratified layer is formed. The size of the experimental tank
allowed visualisation of the wave propagation over a large horizontal distance.
Furthermore, the platform can serve as a
turntable, but the experiments discussed in this paper were done without rotation.

\subsection{Experimental parameters}

A vertically stratified layer of height 80 cm was created by
filling the tank from below with water whose salinity was gradually
increased; {\color {black} this resulted in a layer whose density $\rho$ decreased
from 1040 kg/m$^3$ at the bottom to 1010 kg/m$^3$ at the top. Thus, a
layer of linearly stratified fluid was formed with a constant value $N_0=0.58$ rad/s of the Brunt-V\"ais\"al\"a frequency, 
\begin{equation}
N = \sqrt{-\frac{g}{\rho_0} \frac{d\rho}{dz}}\,,
\end{equation}
also referred to as the buoyancy frequency later on.
The first experiments were carried out with this layer alone.
It must be noted that the buoyancy frequency in this layer remained constant in subsequent days, except for the lowest part of the tank where it became slightly weaker with time, as can be seen in Figure~\ref{all_bv_find_gamma}\,(a), due to molecular diffusion.}

Then, a thin layer (viz.\ 4 cm) of fresh water ($\rho\approx$ 1000
kg/m$^3$) was added from above on top of the uniformly stratified
layer. This served as the upper mixed layer, while the sharp
transition of density between the two layers ensured the presence of a
pycnocline. {\color {black} After a day}, the pycnocline tended to lose its sharpness, and so new fresh layers were added after experiments, or sometimes the upper part of the water column was removed before adding fresh water.
In this way, a whole range of different pycnoclines (in terms of peak value {\color {black}$N_\text{max}$}, thickness, and depth) was created, serving as varying background conditions during
which the experiments were carried out. {\color {black} Examples of 
stratification profiles are provided in Figure \ref{all_bv_find_gamma}~(a), and characteristic values are presented in Table~\ref{tab_params}.}  {\color {black} Density measurements done after each experiment showed no significant changes compared to the profile taken before the experiment.}

{\color {black}The ratios between $\omega$ and the stratification frequencies, $N_0$ and $N_\text{max}$, are both of the order of $10^{-1}$ in our experiments, which is larger than the ones in the ocean. Typical values for the ocean are $N_0=10^{-3}$\,rad\,s$^{-1}$ and $N_\text{max}=10^{-2}$\,rad\,s$^{-1}$, which give for the semi-diurnal tide ($\omega=1.4~10^{-4}$\,rad\,s$^{-1}$) ratios of the order of $10^{-1}$ and $10^{-2}$ respectively. This implies differences in the steepness of the incident internal wave beam, and the experimental results might not be directly applicable to the ocean.}

\subsection{Incident internal wave beam}

\textcolor {black}{In the basin, an internal wave beam directed upwards was generated by a wavemaker similar to the one developed by \cite{gost07a}, to which we refer for a detailed description. As described in \cite{merc10}, the internal-wave beam profile thus obtained is very similar to the solution derived by \cite{thom72}. Their profile has previously been shown to describe wave beams in the ocean after they propagate far from the continental shelf~\citep{gost07b}. In our specific setup, we can directly reproduce such a `tidal' beam without topography and without barotropic flow.}

Briefly, the device consists of $24$ stacked PVC sheets of $2$~cm thickness, shifted with respect to each other so as to form a wave pattern similar to the Thomas-Stevenson wave beam described in \S\,4 in~\cite{merc10}.
The PVC sheets in the wavemaker are connected via two eccentric camshafts, which, when subject to rotation, results in downward (phase) propagation of the wave pattern.
{\color {black} At both ends of the stacking, there are sheets with zero amplitude of oscillation so that when measured along the wavemaker, the width of the generated beam is of the order of $33$~cm.
The amplitude of oscillation of the plate in the center of the beam is $3.5$~cm.}
Placed in a stratified fluid at an angle of $30^\circ$ with respect to the vertical, as depicted in Figure~\ref{scheme_exp}\,(a), this wave motion is imparted to the fluid, thus producing a unidirectional internal wave
beam at a single frequency.
The device had a finite width of 1.5 m along the $y$-direction, resulting in three-dimensional effects close to the edges.

\begin{table}
\label{exp-values}
\begin{center}
\begin{tabular}{c|cccccc|ccccc}
EXP & {\color {black}$N_{\text{max}}$} & $T$ & \textcolor {black}{$U_{\text{max}}$} & $k_\eta$ & $\lambda_x$ & $C_x$ & $\gamma$ & \textcolor {black}{$\omega/N_0$} & \textcolor {black}{$\omega/N_\text{max}$} & \textcolor {black}{$Re$} & \textcolor {black}{$A_k$} \\
$\#$ & (rad\,s$^{-1}$) & (s) & (cm\,s$^{-1}$) & (rad\,m$^{-1}$) & (cm) & (cm\,s$^{-1}$) & & & & &\\
\hline
03 & {\color {black}0.58}  & 21.6 & \textcolor {black}{0.74} & 31.7 & 39.5 & 1.83 & 0 & \textcolor {black}{0.50} & \textcolor {black}{0.50} &\textcolor {black}{232}  & \textcolor {black}{0.80} \\
08 & 2.27 & 21.6 & 0.74  & 31.7 & 39.5 & 1.83 & \,0.11\, & \,0.50\, & \,0.13\, & \,232\, & \,0.80\,\\
16 & 1.52 & 21.6 & 0.74  & 31.7 & 39.5 & 1.83 & 0.19 & 0.50 & 0.19 & 232 & 0.80 \\
19 & 1.94 & 21.6 & 0.74  & 31.7 & 39.5 & 1.83 & 0.24 & 0.50 & 0.15 & 232 & 0.80 \\
21 & 1.83 & 41.8 & 0.40  & 32.8 & 73.8 & 1.77 & 0.17 & 0.26 & 0.08 & 122 & 0.87 \\
22 & 1.83 & 21.6 & 0.74  & 31.7 & 39.5 & 1.83 & 0.17 & 0.50 & 0.16 & 232 & 0.80 \\
25 & 1.83 & 16.9 & 0.68  & 32.2 & 30.4 & 1.80 & 0.17 & 0.64 & 0.20 & 217 & 0.60 \\
\end{tabular}
\end{center}
\caption{Name of the experiment; the maximum buoyancy frequency in the pycnocline $N_\text{max}$; the forcing period $T$; the maximum velocity in the direction of propagation of the initial beam $U_\text{max}$; the dominant wavenumber of the initial beam $k_\eta$; its horizontal wavelength, $\lambda_x=2\pi/k_\eta\sin\theta$, where $\theta$ is the angle of the beam with the horizontal; and finally, the corresponding horizontal phase speed $C_x=\lambda_x/T$. Relevant non-dimensional parameters are also provided: $\gamma$, a measure of the strength of the pycnocline (as explained in section~\ref{subsect:pyc-strength}); the ratio of forcing and stratification frequencies; the Reynolds number $Re=U_{\text{max}}/\nu k_{\eta}$ and the excursion parameter $A_k=U_{\text{max}} k_{\eta}/\omega$.}
\label{tab_params}
\end{table}

Quantitative measurements of the velocity field and the pycnocline displacements were made using Particle Image Velocimetry and ultrasonic probes, respectively.
{\color {black} From the PIV velocity measurements described below, several properties of the beam can be estimated. We measured the maximum amplitude of the velocity $U_{\text{max}}$ in the direction of propagation of the beam, and the dominant wavenumber $k_\eta$ associated with the profile. These properties, along with other relevant parameters in the experiments, are listed in Table~\ref{tab_params}. It is interesting to notice that although our experiments are at quite low values for the Reynolds number, nonlinearity is expected to play an important part since the excursion parameter, \mbox{$A_k=U_{\text{max}} k_{\eta}/\omega$}, which compares the fluid particle displacement with the characteristic wavelength of the problem, is close to $1$.}

\subsection{Particle Imagery Velocimetry (PIV)}

Velocity fields have been obtained through PIV measurements done in
the $(xOz)$ plane as shown in Figure~\ref{scheme_exp}\,(c) (grey dashed line). Two CCD
12-bit cameras of resolution $1024$\,x\,$1024$ record at $3$\,Hz images of
particles illuminated by a $6$\,W continuous laser.  Although the
spatial resolution obtained is good, smaller than $1$~mm/pix, it is
only possible to visualise the linearly stratified region due to the
strong optical distortions caused by the presence of the pycnocline.

Furthermore, the seeding of particles in the homogeneous surface layer
is insufficient since the particles used here rather match the density
range of the stratified layer.  Some information can be extracted from
floating particles at the free surface, as we will see in
section~\ref{sec:meanflow}, which allow us to extrapolate the velocity
field from the the top of the linearly stratified region up to the
free surface.

Experimental data are processed using open source CIVx algorithms by \cite{finc00} and the free Matlab toolbox UVMAT developed at LEGI.
Some overlap between the fields of view of the two cameras allows for a horizontal merging of the velocity fields, 
leading to visualisation windows of $1.6$\,x\,$0.8$\,m.

\subsection{Ultrasonic probes}
\label{ultrasonic_description}

The ultrasonic probes designed at LEGI~\citep{mich97} consisted of a sound emitter, which also acts as a
receiver, and a reflector that were $0.05$m apart. Ten such probes
were positioned at $x=0.18$, $0.39$, $0.62$, $0.89$, $1.20$, $1.254$,
$1.59$, $2.09$, $2.80$ and $4.0$~m from the wave generator, respectively
(see Figure \ref{scheme_exp}\,(a)). Each probe was mounted vertically
on a plane that was slightly off-centered so as to not interfere with
the PIV measurements, at a height that ensured that the emitter was
in the mixed layer and the reflector in the lower constant
stratification layer. The voltage output from the probe is
proportional to the time taken for the sound to traverse its path from
the emitter to the reflector and back to the emitter/receiver. This
traverse time is directly proportional to the relative distance
between the emitter and the pycnocline, if the pycnocline is
infinitesimally sharp. In realistic scenarios, like the stratification
in our experiments, where the pycnocline is of finite thickness, the
output from the ultrasonic probe is a measure of the average (over
$z$) displacement of the pycnocline at the specific $x$ location of
the probe.


To calibrate each ultrasonic probe, it is moved using a linear
traverse mechanism up and down by known distances in a quiescent
setting where there is no flow, and hence no pycnocline
displacements.  The output voltage is continuously recorded as a
function of the position of the probe.  This allows us to get a
voltage $V_p$ vs. relative position of the pycnocline $z_p$
relation for the probe (see Figure~\ref{scheme_exp}~(b)), and the
procedure is done for every probe. Converting the output voltage from
experiments to equivalent pycnocline displacements involves
interpolation on the respective calibration curves.


It is important to note that the values thus found are a measure of
the mean vertical displacements between the emitter and receiver. A
mode trapped in the pycnocline will have maximum vertical excursions
at a certain depth, while the excursions decline rapidly above and
below. The acoustic probes, then, will not yield this maximum
excursion, but a lower value, because it also registers over parts in
which the amplitude is declining.
{\color {black} This reduction will affect the signal associated with even modes (i.e., modenumber 2, 4 etc.), in particular. }

%% file: 3_flow.tex
\section{Mean flow and harmonics}
\label{sec:3}

\begin{figure}
\begin{center}
\includegraphics[width=12.5cm]{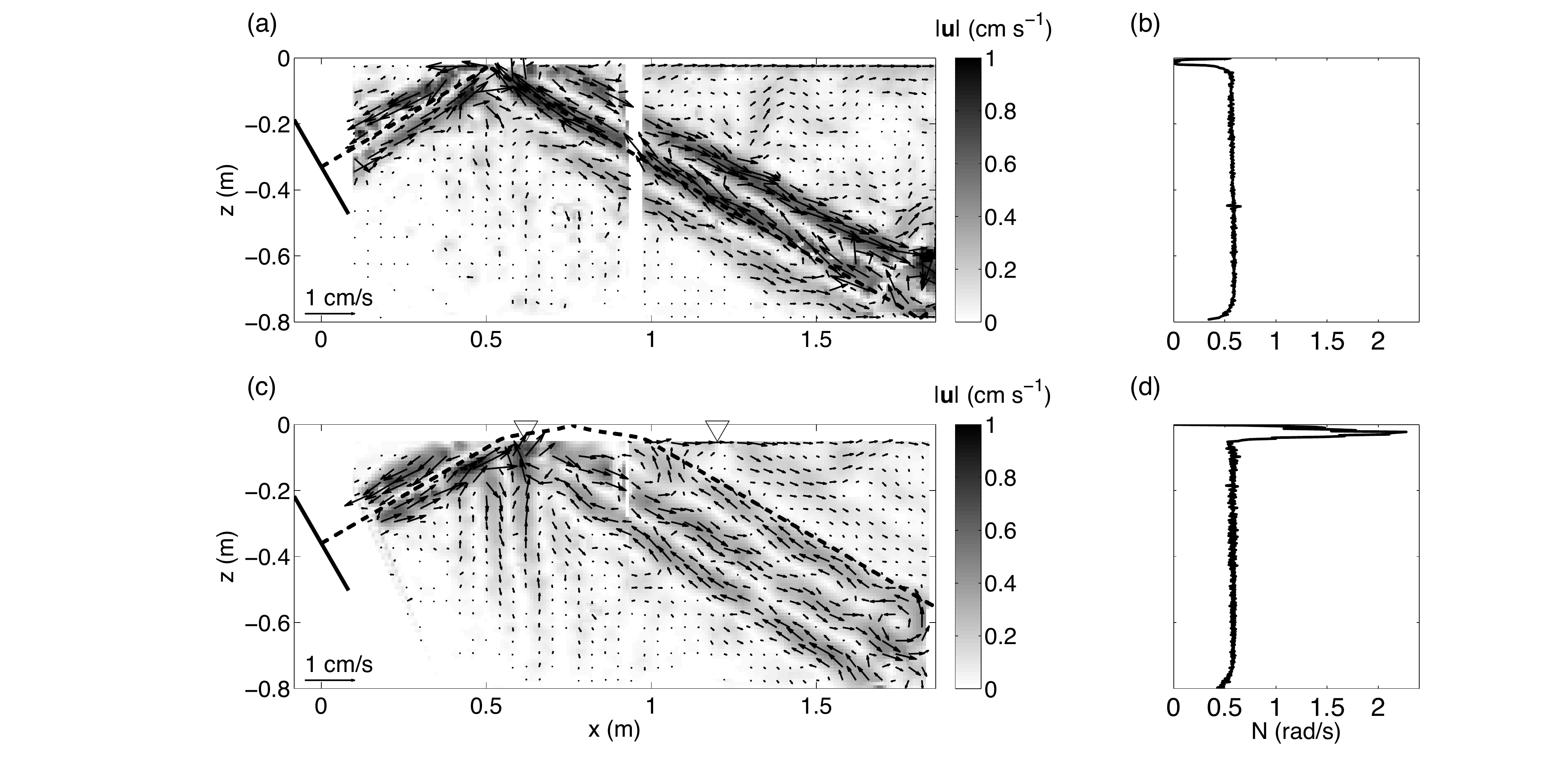}
\caption{Comparison between experiments without (upper panels, EXP03)
and with a pycnocline (lower panels, EXP08). The stratification
profiles are shown in (b) and (d), and the corresponding measured velocity vectors and magnitude 
at $t=240$~s after the start-up of the forcing in (a) and (c), respectively. The dashed lines represent 
the theoretical direction of internal-wave energy propagation, starting from the center of the wavemaker, 
which is indicated by the solid line on the left. Insufficient overlap between the two cameras, 
around $x=0.9$~m, leads to a gap in the velocity field in (a). 
The two inverted triangles in (c) indicate the positions of the ultrasonic probes at $x=0.62$~m and $x=1.20$~m.}
\label{EXP03_08_qualitative}
\end{center}
\end{figure}

Using PIV, we visualise what happens when the incident wave beam
encounters the pycnocline and subsequently reflects from it. To
demonstrate the significance of the presence of a pycnocline, we first
present a qualitative comparison between settings without and with a
pycnocline, see Figure\ \ref{EXP03_08_qualitative}. The effect of the
pycnocline, as seen in the unfiltered PIV data presented in the lower
panel of Figure\ \ref{EXP03_08_qualitative}, is twofold. First, it
shifts the beam slightly to the right because the beam propagates at a
smaller angle with respect to the horizontal in the layer of strong
stratification (as indicated by the dashed line which represents the
theoretical path of energy propagation).
Second, the reflected beam is broadened, and therefore less intense,
when it emerges from the pycnocline. This is caused by (multiple)
internal reflections: the beam reflects not only from the upper free
surface but also from and within the pycnocline itself \citep{gerk01,math09}.

As mentioned in the introduction, two kinds of nonlinear phenomena are
expected to take place during reflection at the pycnocline: the
nonlinear evolution of the pycnocline displacements excited by the
incoming beam, and the generation of higher harmonics at the junction
between the incident and reflected beams. The latter occurs even if no
pycnocline is present, namely upon reflection from the surface. As
noted earlier, the upper most part of the water column produces no
visible signal with PIV in the experiments with a pycnocline.
Therefore, we resort to the acoustic probes for visualising and analysing
the response at the pycnocline, the subject of section 4.
In this section, we focus on the features extracted from the PIV data, notably the
principal beam and its harmonics.

The harmonics cannot be properly interpreted unless we take
into account another phenomenon that we found in our experiments: the
occurrence of a mean flow in the upper layer. We therefore proceed to
discuss this phenomenon first.

\subsection{Mean flow}
\label{sec:meanflow}

\begin{figure}
\begin{center}
\includegraphics[width=12.5cm]{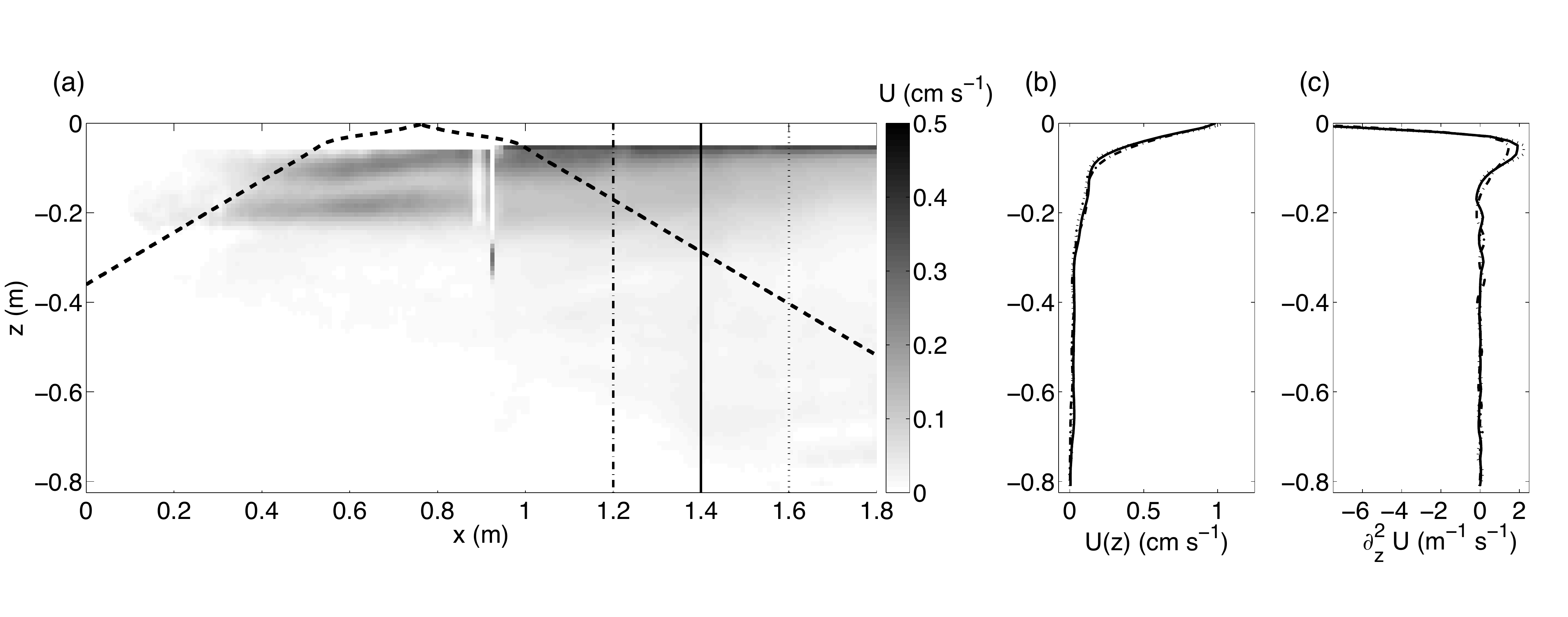}
\caption{EXP08, (a) Mean horizontal velocity $U$ and vertical profiles of (b) $U$ and (c) $\partial^2U/\partial z^2$ extracted at $x=1.2$~m (dashed-dotted line), 
$1.4$~m (continuous line) and $1.6$~m (dotted line) from the wavemaker. 
The thick dashed line represents the theoretical direction of internal-wave energy
propagation, starting from the center of the wavemaker. Insufficient
overlap between the two cameras, around $x=0.9$~m, leads to a gap in the
velocity field.}
\label{EXP08_meanflow_shear}
\end{center}
\end{figure}

Apart from extracting the higher harmonics from the PIV data, harmonic
analysis also reveals the presence of a mean {\color {black} Eulerian} flow. The mean flow for
EXP08, in which the cameras were placed close to the wavemaker, is
shown in Figure \ref{EXP08_meanflow_shear}.  It is found to originate
from the area of reflection of the forced beam at the pycnocline, and
{\color {black} to} spread further in the direction of wave propagation, 
but {\color {black} it} 
remains restricted to the upper layer. The horizontal mean velocity displayed
here is obtained by averaging over 12 forcing periods, starting at
$t=360$~s after the start-up of the forcing.

A complete description of the mean-flow field is not directly
available as the PIV data lacks information in the upper few
centimetres, typically 5 cm, with the exact value depending on the
experiment. This is partially remedied by determining, from the raw
PIV images, the strength of the mean flow at the free surface ($z=0$) by
following the particles trapped there by capillary effects.  This
allows us to establish the profile of the mean flow over the entire
vertical, without having to extrapolate; we only interpolate
(cubically) over the interval where data is lacking.
{\color {black} Furthermore, we also impose the constraint of no shear at the free surface during interpolation.}
Figure\,\ref{EXP08_meanflow_shear}\,(a) and (b) illustrate {\color {black} 
the resulting} profiles.
The second vertical derivative of the mean horizontal velocity, plotted in Figure\,\ref{EXP08_meanflow_shear}\,(c), is an important quantity as it features in equation~(\ref{modes-shear-W}), determining
the vertical modal structure.  A noticeable characteristic of the mean flow
is that its amplitude varies with $x$, being maximum where it
originates (at the area of reflection of the forced beam at the
pycnocline) and then decaying slightly as you go away from the
wavemaker.
This evolution must be a signature
of a three-dimensional structure of the flow; the amplitudes of the
wave field at the forcing and higher harmonic frequencies, however,
did not exhibit a similar decay in $x$. Hence, for the present
purposes we assume the flow to be quasi-two-dimensional.

\begin{figure}
\begin{center}
\includegraphics[width=0.5\textwidth]{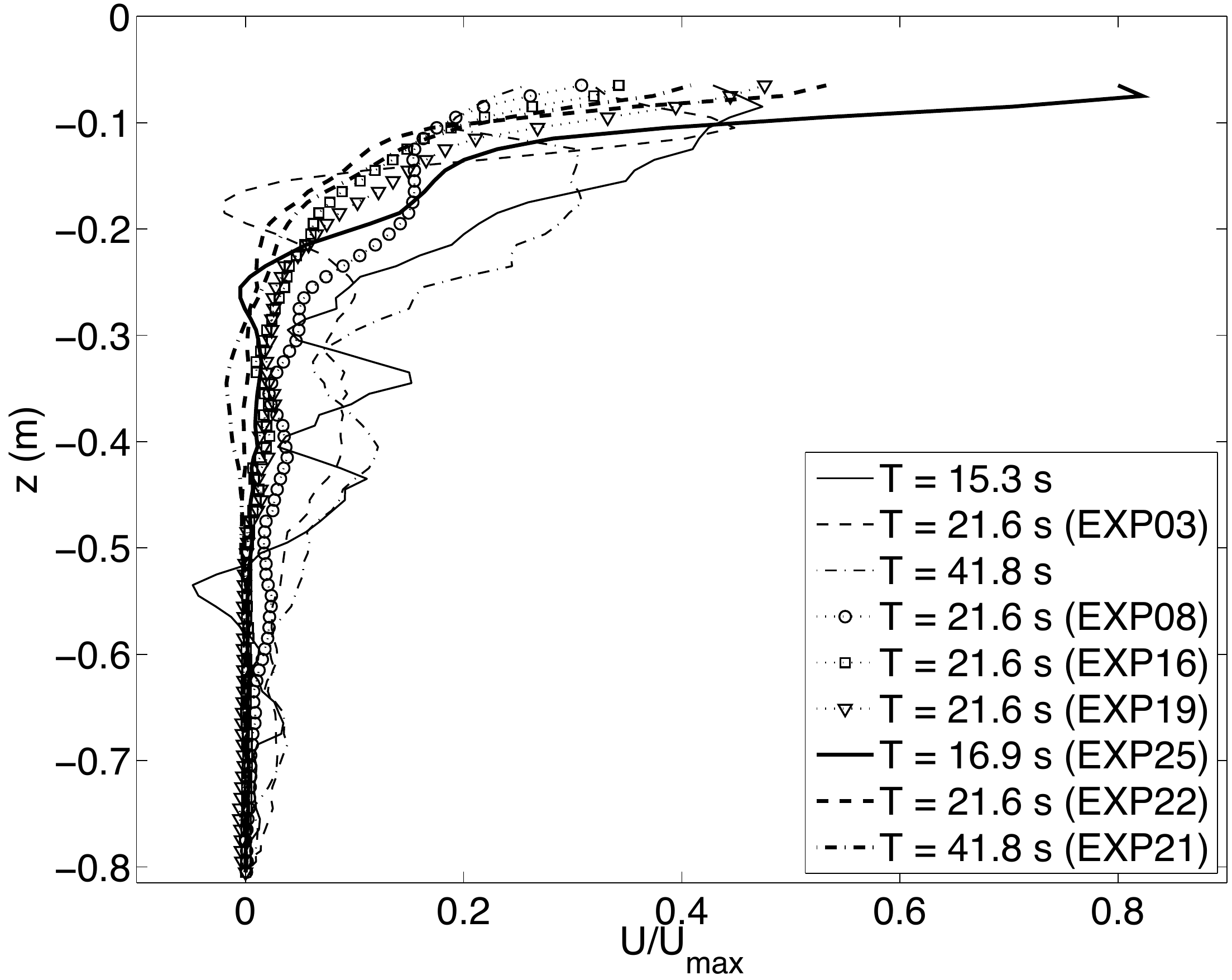}
\caption{{\color {black}Vertical profiles of the mean horizontal velocity $U(z)$ extracted  at $0.4$~m away from the point of the impingement of the incident beam on the free surface, for various experiments. The two profiles (thin lines) not associated with any experiments are with the same pycnocline strength as EXP03.}}
\label{ALLEXP_meanflow}
\end{center}
\end{figure}

{\color {black}Interestingly, the mean flow is present in all our experiments, including the ones without a pycnocline.
In Figure~\ref{ALLEXP_meanflow}, we plot the vertical profile of the mean flow extracted at $0.4$\,m away from the point of impingement of the incident beam on the free surface for all experiments presented in Table~\ref{tab_params}, and two extra for different frequencies in the cases with $\gamma=0$.
The mean flow maximum velocity is of the same order than the maximum velocity of the incoming wave beam, but no clear effect of the varying pycnocline strength can be observed.
So, it is the reflection from the surface, rather than the passage through the pycnocline, which lies at the origin of the mean flow. Furthermore, we verified that the mean flow is not present at the beginning of the experiment and is noticeable only after the beam impinges on the pycnocline and surface.}

In none of the experiments is an appreciable mean flow found near the
bottom of the tank, even though there, too, beam reflection takes
place. {\color {black}However, since PIV measurements are not possible in the bottom $5$~mm of the fluid, we cannot determine the flow in the vicinity of the wave-induced bottom boundary-layer of thickness $\sqrt{\nu/\omega}$, of the order of $2$~mm in our case.
One could argue that nonlinear effects are stronger at the pycnocline due to the strengthening of 
the wave beam in regions of stronger stratification, and this may result in a net deposit of momentum.
In Figures \ref{EXP03_08_qualitative}\,(a)\,\&\,(c), the first reflected beam coming from the pycnocline/free surface, when arriving at the tank bottom, is not much weaker than the incident forced beam. 
Hence, the contrast between the presence of a mean flow near the surface and its absence near the bottom is presumably due to the different nature of the boundaries, a free surface versus a solid boundary (i.e., free-slip versus no-slip).
These explanations, however, must remain hypotheses since our experiments do not allow us to draw any firm conclusions.
As a matter of fact, mean flow generation near a solid boundary has already been observed in experiments and numerical simulations of barotropic flow over a three-dimensional topography~\citep{king09}. Nevertheless, the mechanism at play in our case is different since we do not have any topography, there is no barotropic forcing, and the mean flow generated is in the same plane as the incoming wave beam.}

Other experiments dedicated to this mean flow are needed to better
understand its cause and its possibly three dimensional structure, as
well as its dependence on various parameters (impinging beam
angle/profile, strength of the pycnocline etc.). As discussed in the
following sections, incorporating the effects of the mean flow plays a
crucial role in the interpretation of the higher harmonics and other
nonlinear waves in our experiments.

\subsection{Higher harmonics}

Here, we present a harmonic analysis of the PIV data from three
experiments that were carried out on the same day, hence {\color {black} with nearly identical density profiles (all corresponding to $\gamma=0.17$)}, but at different forcing frequencies:
$\omega_f=0.15$ rad/s (EXP21), $\omega_f=0.29$ rad/s (EXP22),
$\omega_f=0.37$ rad/s (EXP25). These three experiments had the cameras
placed further away from the wavemaker than in EXP08, and hence
provide a more complete view of the region after reflection from the
pycnocline.

\begin{figure}
\begin{center}
\includegraphics[width=12.5cm]{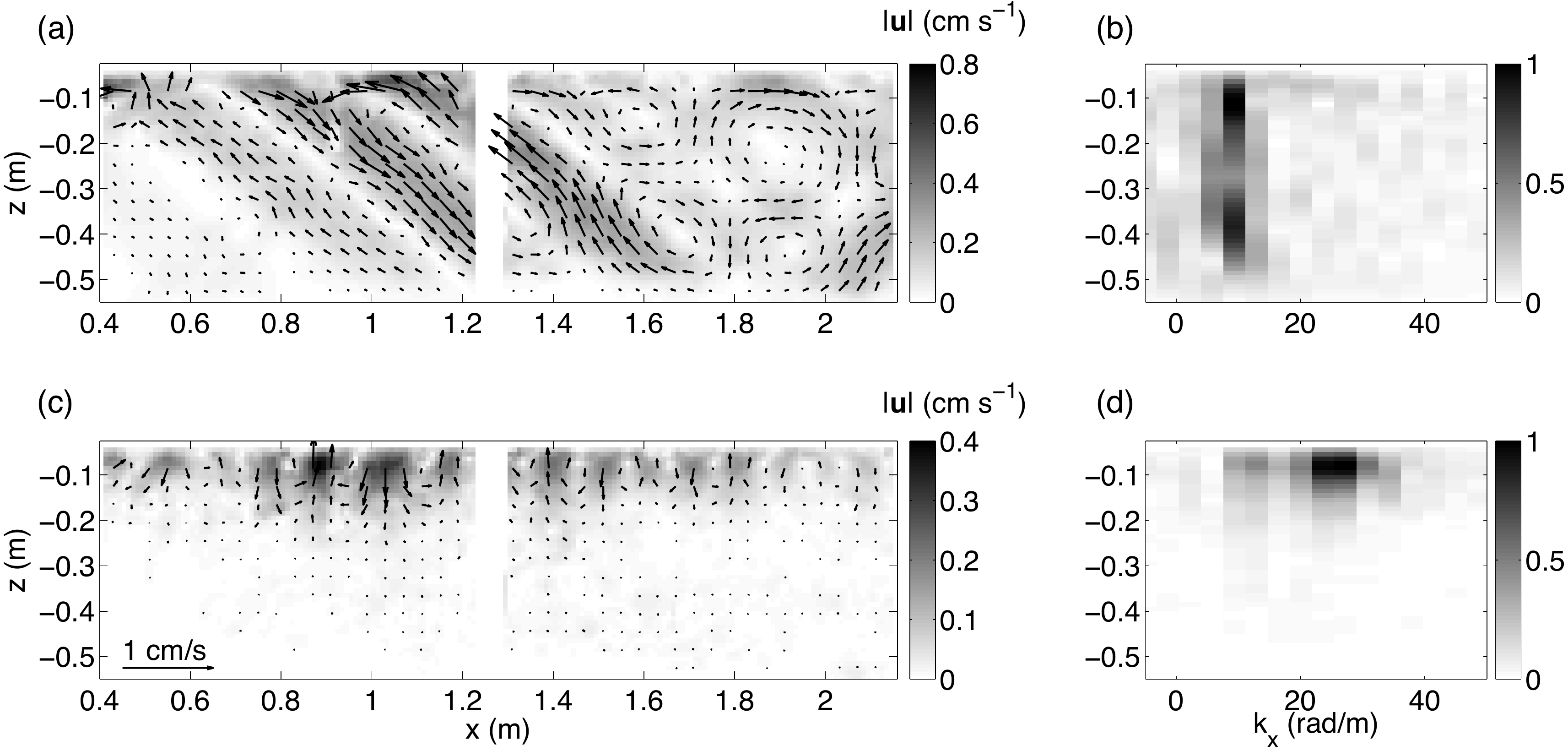}
\caption{EXP25, harmonic analysis for the forcing frequency $\omega_f$ (upper panels), and the first harmonic $2\omega_f$ (lower panels). Panels (a) and (c) display the velocity vector fields and its amplitude. The arrow indicating the scale of the velocity vectors in (c) is also valid for (a). Panels (b) and (d) show the horizontal Fourier spectra of the vertical velocity $w$ as a function of the horizontal wavenumber $k_x$ and its vertical variation (for each, {\color {black} amplitudes are normalised by the maximum value)}.}
\label{EXP25_harmonics}
\end{center}
\end{figure}

In EXP25 {\color {black}($\omega_f/N_0=0.64$)}, presented in Figure~\ref{EXP25_harmonics}, the principal
beam at the forcing frequency reflects from the pycnocline and
surface, and broadens due to internal reflections. The dominant
wavenumber in the reflected beam, $k_x^{(r)}\simeq10.7$~rad/m is much
smaller than the corresponding value in the incident wave beam,
$k_x^{(i)}=20.7$~rad/m.  The reflected beam at the forcing frequency,
which radiates away from the pycnocline into the lower layer, is in
stark contrast with the higher harmonics $2\omega_f$ (and $3\omega_f$,
not shown in the figure), which are trapped in the upper part of the
water column and propagate purely horizontally with a well-defined
horizontal periodicity. For the $2\omega_f$-signal, we find a
horizontal wavelength of 22 cm.  The fact that the harmonics are
trapped is readily understood from linear internal-wave theory, since
their frequencies $2\omega_f,3\omega_f$ etc.\ exceed the buoyancy
frequency of the lower layer ($N_0$); hence the harmonics cannot
propagate into it and are constrained to the pycnocline.

The experimentally measured horizontal wavelength of the trapped
higher harmonics may be compared with the theoretical estimates from
the eigenvalue problem for the vertical modal structure
\begin{equation}\label{modes-shear-W}
W'' + \Big[ \frac{N^2}{(U-c)^2}-\frac{U''}{U-c}-k^2\Big]W = 0 \,,
\end{equation}
with the vertical velocity $w(x,z,t)=\Re[W(z)\exp[i(kx-\omega t)]]$,
and the boundary condition $W=0$ at the surface and bottom~\cite[][Eq.~41.8]{lebl78}.  Here, $c=\omega/k$ is the horizontal phase
speed, $U(z)$ is the background mean flow profile, $U''(z)$ its second derivative and $N(z)$ the
buoyancy frequency profile. We note that this equation is different
from the one in terms of the vertical isopycnal displacement, $\eta$.
In the literature, the two forms are sometimes confused. The
relation between the two follows from $w=\eta_t+U\eta_x$, so that
$W=ik(U-c)\phi$, where $\eta=\Re\{\phi(z)\exp[i(kx-\omega t)]\}$.

\begin{figure}
\begin{center}
\includegraphics[width=0.5\textwidth]{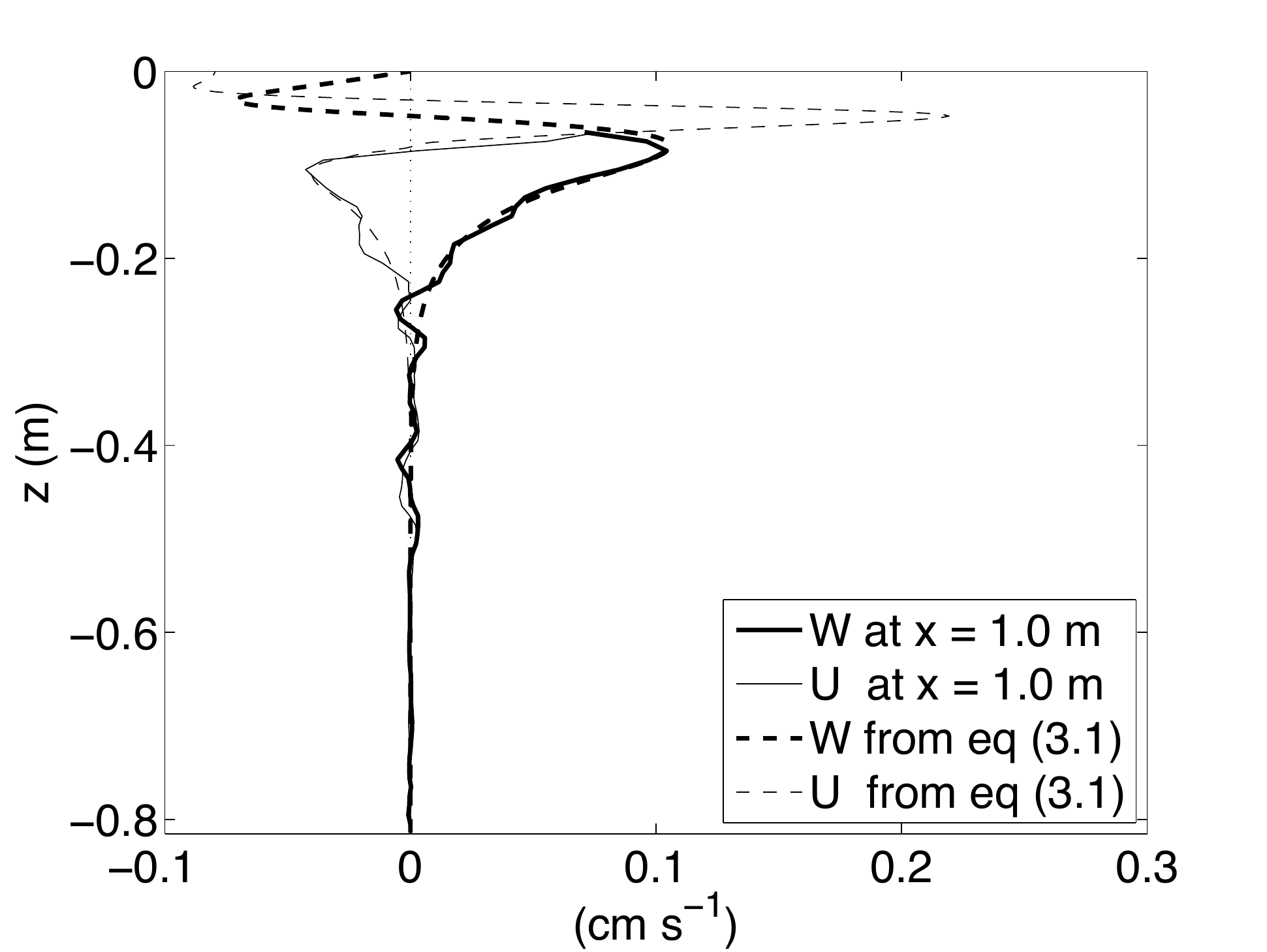}
\vspace{0.01cm}
\caption{EXP25, Comparison between the mode-2 shape from
experimental velocity profiles at $x=1.0$~m, filtered at $\omega=2\omega_f$, and the ones calculated from (\ref{modes-shear-W}).}
\label{EXP25_comparison}
\end{center}
\end{figure}

For $\omega=2\omega_{f}$, and observed $N(z)$ and $U(z)$, we solve
(\ref{modes-shear-W}) to obtain the modes $W_n$ and corresponding
wavenumbers $k_n$ or wavelengths $\lambda_n=2\pi/k_n$. For EXP25, we
find {\color {black}$\lambda_1=48.9$, $\lambda_2=20.2$ and $\lambda_3=14.8$ cm}, 
using the background mean flow profile measured at $x=1.0$m.
The observed horizontal wavelength (i.e., 22cm) of the signal at $\omega=2\omega_f$
is the closest to that of the theoretical 2$^{nd}$ mode. This is confirmed by the excellent agreement between the
experimental and theoretical profiles of $W_2$ and $U_2$, as shown in
Figure \ref{EXP25_comparison}, at least so far as the comparison can
be made (as mentioned earlier, PIV data is not available in the upper
few cm).
Exactly why the first harmonic (i.e.,\ $2\omega_f$) manifests itself as a second mode, and not as some other mode, is not clear.
{\color {black}For instance, we note that the horizontal phase speed of the incident internal wave beam, $1.8$~cm~s$^{-1}$, is not particularly close to either the measured ($2.6$~cm~s$^{-1}$) or theoretically calculated ($2.4$~cm~s$^{-1}$) mode-2 phase speed for $\omega=2\omega_f$.
In fact, it is closest to the theoretically calculated horizontal mode-3 phase speed for $\omega=2\omega_f$ ($c_3=2\omega_f/k_3=1.8$~cm~s$^{-1}$).
This suggests that a phase speed matching mechanism, such as the one discussed by \cite{gris11}, who examined the local generation of internal solitary waves with a numerical model but without observing any mean flow, is not in play here.
Two possible explanations may be advanced.
In a weakly nonlinear case where there is just advection of the harmonics by the mean flow, we could search for a criterion with Doppler effect on the phase speed of the harmonics. Another more complex scenario corresponds to fully nonlinear coupling of the mean flow and the harmonics. However our experiments cannot provide an answer to this point of high interest.}

{\color {black}Finally,} the importance of the shear flow to the modal propagation of the
higher harmonics can be illustrated by solving the modes from
(\ref{modes-shear-W}) with $U=0$; then, for $\omega=2\omega_f$, we
find much smaller horizontal wavelengths: $\lambda_1=41$,
$\lambda_2=13$ and $\lambda_3=8$~cm, none of which matches the
observed wavelength.

For EXP22 {\color {black}($\omega_f/N_0=0.50$)}, we again find that the measured horizontal wavelength ($35$ cm) of the harmonic $2\omega_f$ is closest to the corresponding
theoretical wavelength of mode-2 
{\color {black}($\lambda_1=74$, $\lambda_2=29$ and $\lambda_3=20$ cm).}
We note, however, that the wavelengths are in this case ($2\omega_f\approx N_0$) sensitively dependent on the frequency $\omega=2\omega_f$ and $N_0$, because they are very close. For example, the mode-2 wavelength for $\omega=0.5$ and $0.6$~rad/s for the stratification in EXP22 are $\lambda_2=73$ and $24.5$~cm, respectively. Thus the quantitative agreement
between experiment and theory could be better or worse for even small
changes in $\omega$ or $N_0$. As it was the case for EXP25, the vertical
structure of the first harmonic signal ($2\omega_f$) agrees well with
the theoretically computed mode-2 shape, except in the deep parts of
the lower layer. In the lower layer, where the measured background
shear flow is negligible, equation (\ref{modes-shear-W}) predicts a
linear variation of $W$ with $z$ for $\omega=2\omega_f\approx N_0$
(since the equations boils down to $W''\sim0$); however, we observe a
more rapid decay of the signal in the lower layer, the cause of which
remains unclear.

\begin{figure}
\begin{center}
\includegraphics[width=12.5cm]{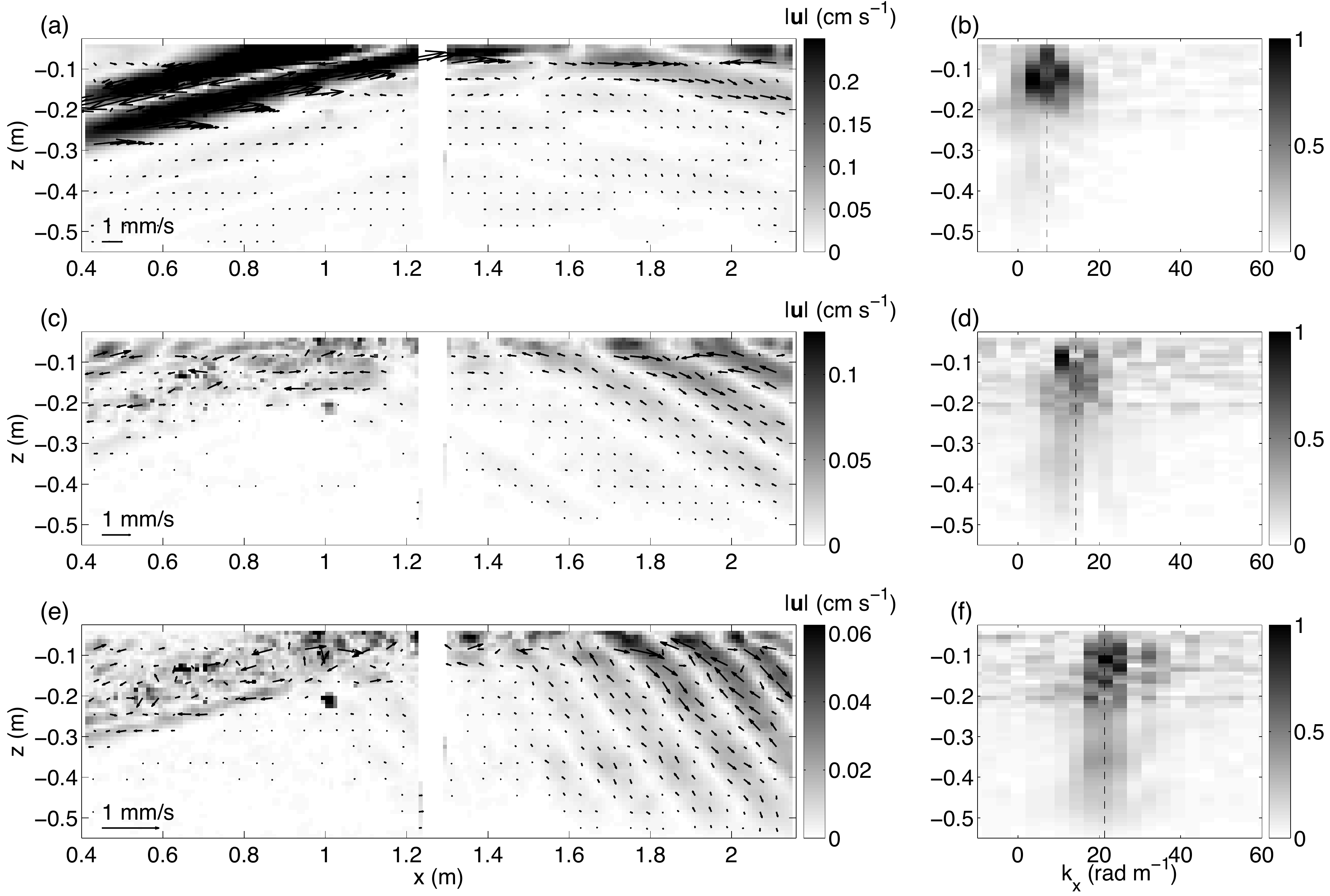}
\caption{EXP21, harmonics analysis as in Figure\,\ref{EXP25_harmonics}
for the frequency $n\omega_f$, with $n={1,2,3}$. The dashed lines in (b), (d) and (f) represent, for
harmonic $\omega=n\omega_f$, the value $k^{(n)}=nk^{(1)}$ with
$k^{(1)}=7.1$~rad~m$^{-1}$.}
\label{EXP21_harmonics}
\end{center}
\end{figure}

In the case of EXP21 {\color {black}($\omega_f/N_0=0.26$)}, for which $3\omega_f<N_0$, the first two
harmonics are not trapped but are allowed to propagate into the lower
layer; hence they radiate from the pycnocline as beams and cannot be
identified with any one mode, as shown in
Figure\,\ref{EXP21_harmonics}. However, a different simple regularity
appears from Figure\,\ref{EXP21_harmonics}. For each harmonic ($n$), the
beam pattern is clearly periodic (see panels on the left); the
horizontal wavenumber associated with this periodicity ($k^{(n)}$,
say) obeys the simple rule $k^{(n)}=n k^{(1)}$ (see dashed lines in
panels on the right of Figure\,\ref{EXP21_harmonics}).  This implies that
the horizontal phase speeds are the same for all harmonic beams.

Finally, it is evident from Figure\,\ref{EXP25_harmonics} and
Figure\,\ref{EXP21_harmonics} that the amplitude of the harmonics at
$n\omega_f$ is weaker than the one at $(n-1)\omega_f$ (for $n>1$), and
that these higher harmonics originate from the region of interaction
between the incident wave beam and the pycnocline. These observations
confirm that the higher harmonics are a result of weakly nonlinear
interactions between the incoming and reflected beams. 
{\color {black} Since the Reynolds number varies along with other parameters through the experiments ({\it cf.} Table~\ref{tab_params}), we are unable to discriminate its role on the generation of harmonics, more specifically its effect on the saturation of the harmonics amplitude as already observed in~\cite{king09} for instance.
Varying the forcing frequency across various experiments results in the value of $\omega_f/N_0$ go from values smaller to larger than 0.5, and thus influencing the propagation in the lower constant stratification layer of higher harmonics; freely propagating for $\omega_f/N_0<0.5$ and trapped in the pycnocline for $\omega_f/N_0>0.5$. The change in the ratio $\omega_f/N_\text{max}$ however, remains weak and does not seem to play any role in our experiments.}
A detailed study of the influence of the amplitude of the incoming beam on the
amplitude of the generated harmonics is recommended for future
research.

%% file: 4_pycnocline.tex
\section{Evolution of the pycnocline}
\label{sec:4}

In this section, we discuss the evolution of the pycnocline, as
measured by the acoustic probes. As mentioned in Section \ref{ultrasonic_description},
ten acoustic probes were positioned at various positions in the range $0$ to $4$~m
from the wave generator. The proximity of two probes located at $x=1.20$ \&
$1.254$~m, respectively, allows us to determine the horizontal phase speed of the
signal, for its structure remains similar between the two, so that
specific peaks can be identified in both.  In every experiment, the
acoustic probe measurements were initiated well before the wave
generator was started, allowing us to visualise the transients before
a steady state was reached.


\begin{figure}
\begin{center}
\includegraphics[width=12cm]{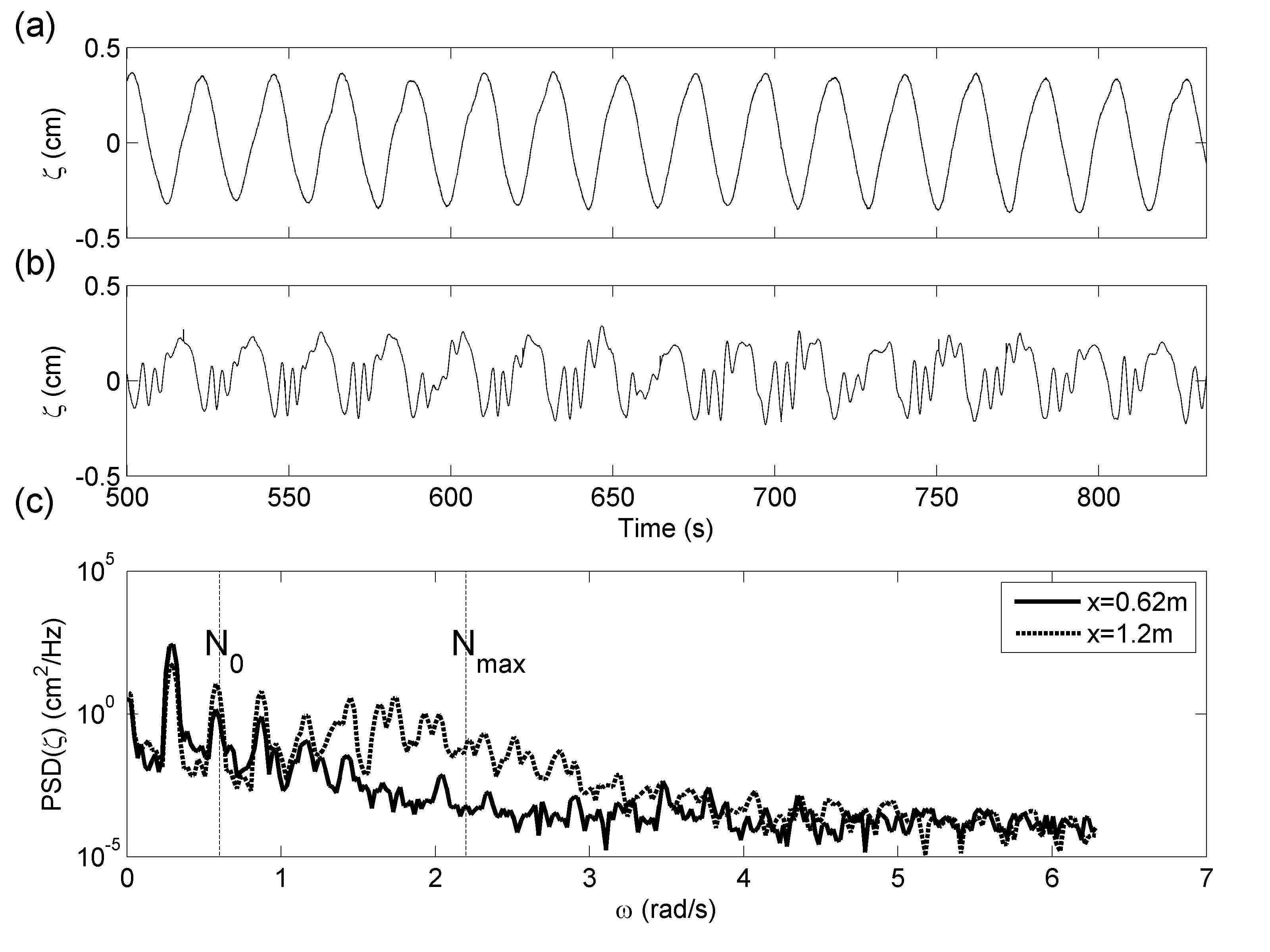}
\end{center}
\caption{EXP08 ($\gamma=0.11$), steady state pycnocline displacements at (a) $x=0.62$~m and
(b) $x=1.2$m. (c) Corresponding Fourier spectra.}
\label{EXP08_pycno_evol}
\end{figure}

In figures \ref{EXP08_pycno_evol}~(a)~\&~(b), we plot the pycnocline
displacements $\zeta$ (i.e., vertical averages over the thickness of the
pycnocline, as described in section \ref{ultrasonic_description}) at $x=0.62$~m and
$x=1.2$~m, respectively.
In a qualitative way, the results from the probes
indicate the formation of ISWs in the pycnocline:
the wave pattern is nearly sinusoidal at the region of impact ($x=0.62$~m),
steepens as it propagates away, and at some point its steep front
splits up into higher-frequency peaks, which are, at least initially ($x=1.2$~m),
ordered in amplitude, the larger ones being ahead.
Furthermore, as shown in figure \ref{EXP08_pycno_evol}~(c), the power spectrum of the
time series at $x=1.2$~m has a distinct bulge around the frequencies
in the neighbourhood of $\omega=N_{max}$, a feature we discuss in
detail later in the current section. In a quantitative sense, however,
it is not obvious that nonlinear effects are at work; after all, the
amplitude of the peaks is small (a few mm) compared to the depth of
the pycnocline (a few cm).

We note, however, that the actual peak amplitudes of interfacial
displacements must be larger than the measured outputs from the ultrasonic
probes as the probes measure the {\em mean} amplitudes over the layer
between the transmitter and the receiver. This effect is particularly
pronounced for the high-frequency waves that are trapped in the
pycnocline, for their amplitude diminishes rapidly outside of the
pycnocline. We also note that the shape and amplitude of the ISWs we find are similar to
those observed in earlier experiments on internal solitons in a
two-layer system (Horn {\it et al.} 2001).

\subsection{Interpretation of spectra}

The generation of higher harmonics due to nonlinear interaction at the
junction of two beams was discussed in Section 3.2. Being relatively
weak, their behaviour must be close to linear as they propagate away
from their source. This means that they can be properly identified in
spectra, as seen in Figure\,\ref{EXP08_pycno_evol}~(c), which after all amounts to treating the signal as if it were linear. For ISWs it is less clear how they might manifest themselves in spectra. Being genuinely nonlinear waves, they are not associated with any particular harmonic; neither can they be described as a superposition of {\color {black} independently propagating} harmonics. Still, technically, one can calculate the spectrum of a signal containing ISWs. As a test case, we use results from a set of KdV-type equations for interfacial waves, which include forcing terms due to barotropic tidal flow over topography \citep{gerk96}.{\color {black} The model is weakly non-hydrostatic and weakly nonlinear; besides, it contains terms describing the advection of baroclinic velocity fields by the barotropic flow}. In one model calculation, we switched off the nonlinear terms while {\color {black} retaining the barotropic advection} terms. 
In this case, the internal-wave field is linear but contains higher harmonics due to the barotropic advection terms. In another calculation, the nonlinear terms were included, giving rise to internal solitons. In this case, there are higher harmonics as well as ISWs, and so this is qualitatively the situation we supposedly have in our laboratory experiments.

\begin{figure}
\begin{center}
\includegraphics[width=12cm]{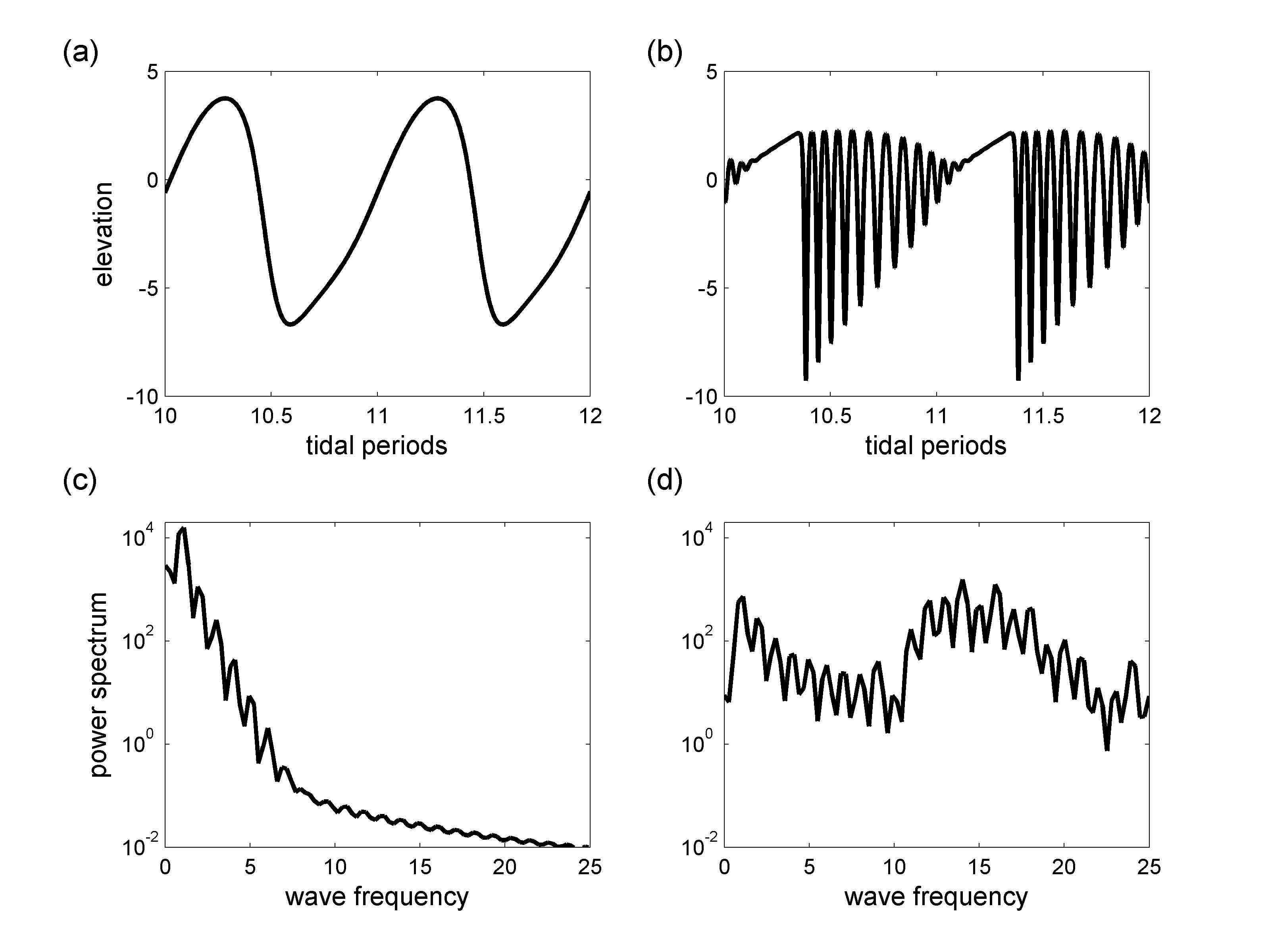}
\end{center}
\caption{Examples from a nonlinear non-hydrostatic KdV
type model for interfacial tidally generated waves. Panel (a) shows the wave profile
over two tidal periods, created from the quasi-linear version of the model,
where higher harmonics are generated by barotropic advection of the baroclinic field;
these harmonics manifest themselves as peaks in the corresponding power spectrum, 
shown in (c). In (b), genuinely nonlinear effects of the baroclinic field itself 
are included, giving rise to solitons; the corresponding spectrum in (d) is now
markedly different, showing a bulge enveloping higher frequencies.}
\label{solspec}
\end{figure}

If we now calculate the spectra for the results of both model runs, we
find important and qualitative differences (see Figure\,\ref{solspec}).
For the quasi-linear case, peaks at higher harmonics are found, and,
as expected, they get lower and lower for higher frequencies.  In the
nonlinear case, the same qualitative behaviour is found at the lowest
frequencies, but then a bulge appears, which is not associated with
any one harmonic, but more wide-spread, involving a group of
harmonics. This is easily understood from the fact that solitons are
not associated with harmonics {\it per se}, but have nonetheless
certain time scales within the range of some of the harmonics. Harmonics in the neighbourhood of these time
scales show marked peaks in the spectrum. This does not mean that the
solitons are formed by a superposition of {\color {black} independently propagating} higher harmonics (after all,
the signal is nonlinear), but are the result of applying a linear
technique (harmonic analysis) to a nonlinear signal. In any case, the
presence of a bulge helps to distinguish the physically distinct cases
of pure higher harmonics on the one hand, and solitons on the other.

Such a bulge appears in our experiments too, as is illustrated in
figure \ref{EXP08_pycno_evol}~(c).  This confirms the direct visual
impression from the probe results shown in figure
\ref{EXP08_pycno_evol}~(b), in which groups of peaked waves are clearly
visible. We proceed, in the next section, to investigate the effect of
the pycnocline properties (depth, vertical extent, strength etc.) on
the bulge and hence propose a way to capture all the properties into
one parameter, the magnitude of which determines the appearance of
solitons.

\subsection{Effect of the pycnocline strength}\label{subsect:pyc-strength}

\begin{figure}
\begin{center}
\includegraphics[width=12.5cm]{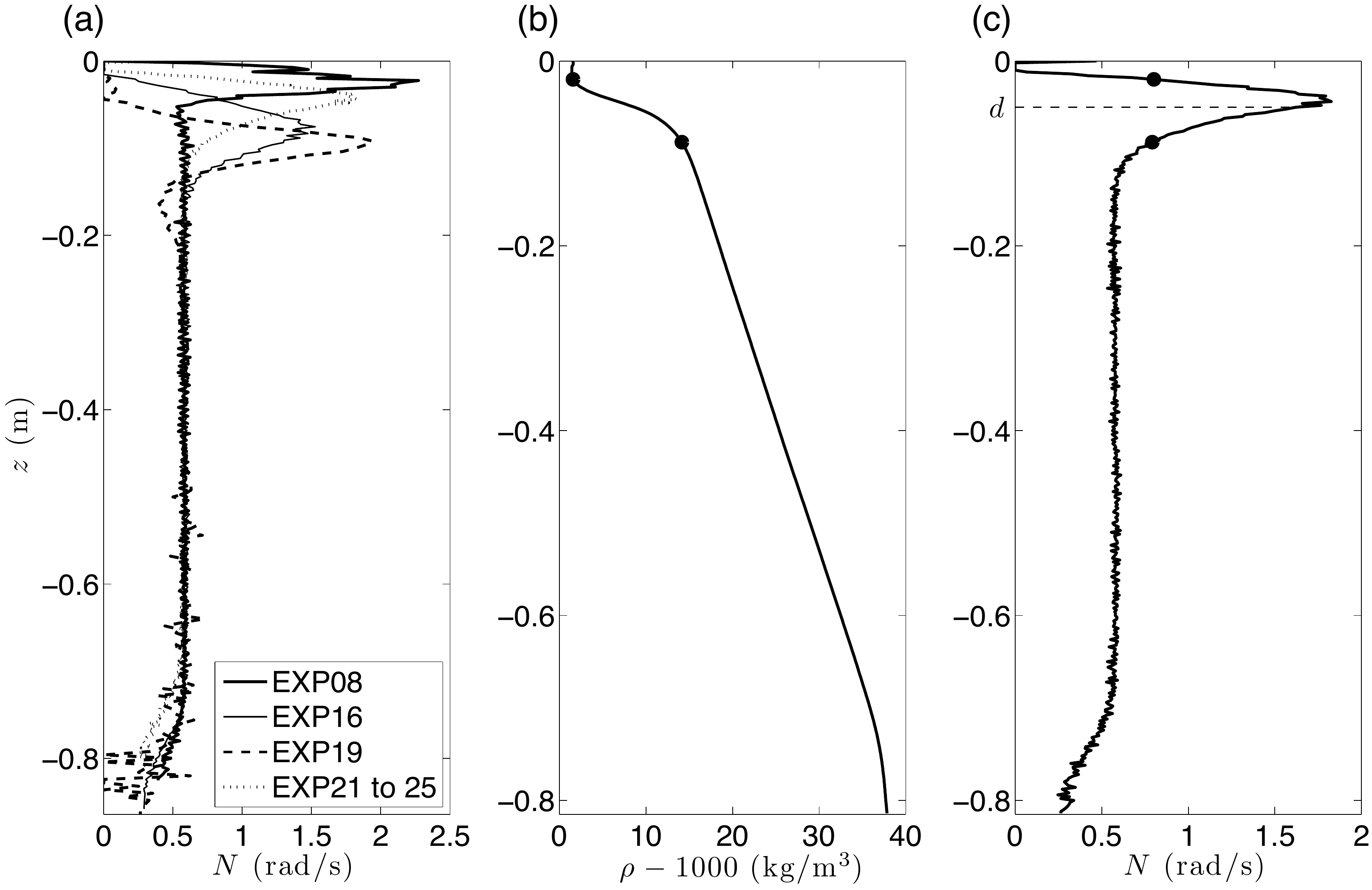}
\end{center}
\caption{(a) Different profiles of stratification with which experiments
were carried out. (b)~\&~(c) Determination of $\gamma$ from the profile for experiments EXP21 to 25. {\color {black}The black dots indicate the (approximate) positions of the top and base of the pycnocline, delineating the vertical interval where density changes strongly in (b)}. The depth of the weighted center of the pycnocline, {\it i.e.} $d$, is indicated by the horizontal dotted line in (c).}
\label{all_bv_find_gamma}
\end{figure}

By removing the upper layer, and refilling it with fresh water, we
varied the depth, the amplitude, and the width of the peak of the
pycnocline (see Figure \ref{all_bv_find_gamma}~(a) for the various stratification
profiles that were set up during the course of all our
experiments). In this section, we present results from three
experiments with the same forcing frequency (such that the incident
internal wave beam propagates at $30^{\circ}$ with respect to the
horizontal) but with different pycnocline properties.

A key parameter for the response of the pycnocline, due to an
impinging internal-wave beam, was identified by Gerkema (2001) using
linear theory. In that study, a ``2c-layer'' stratification was
adopted, consisting of an upper mixed layer of thickness $d$, an
interfacial pycnocline of strength $g'$, beneath which lies a layer of
constant buoyancy frequency, $N_c$. The total water depth being $H$,
the parameter in question is defined by
\begin{equation}\label{gamma}
 \gamma = \frac{(g' d)^{1/2}}{N_cH}\,.
\end{equation}
In the numerator of equation~\ref{gamma}, we recognise the phase speed of long interfacial waves in a 2-layer system,
while the expression in the denominator is a measure of the phase
speed of vertical modes in a constantly stratified layer of depth $H$. These
modes add up to form beams, so the parameter $\gamma$ reflects the
ratio of the phase speeds associated with the interface (i.e.\
pycnocline) and the uniformly stratified lower layer. It was found
by \cite{gerk01} that for $\gamma$ either very small or very large, there is no
significant transfer of energy from beams to interfacial waves. Such a
transfer occurs only for intermediate values, characterised by
$\gamma\approx0.1$. This would therefore be expected to serve as a
necessary condition for having locally generated ISWs which emerge from the initial interfacial perturbation as it
propagates and steepens. It is, however, not a sufficient condition, for the forcing frequency, 
which does not feature in (\ref{gamma}), must also affect the evolution into ISWs.

In the present situation we have a more complicated profile of
stratification, but the parameters in (\ref{gamma}) are still
meaningful if interpreted in the following way.  First, we look at the
density profile to identify the base and top of the pycnocline; an
example is shown in figures \ref{all_bv_find_gamma}~(b)~\&~(c). As explained in \cite{gerk01}, the parameter $g'$ equals the area enclosed by the pycnocline
in the $N^2$ versus $z$ graph; so, $g'=\int N^2dz$, where the
integral is taken between the previously identified base and top of
the pycnocline. The parameter $d$ corresponds to the depth of the
pycnocline, but this depth is of course not uniquely defined here. To
take the depth of the peak would not always make sense because it may
not be representative of where the bulk of the pycnocline
lies. Therefore, we calculated a `weighted depth', as $d=\int z
N^2dz/(\int N^2dz)$, where the boundaries of the integrals are again
the base and top of the pycnocline. The depth $d$ thus found is
illustrated in Figure~\ref{all_bv_find_gamma}~(c). Finally, the
parameter $H$ is simply the total water depth, and $N_c$ is taken
equal to the mean of $N$ from the bottom to the base of the
pycnocline. We have thus obtained all the parameters needed to calculate
$\gamma$, which is listed for different profiles in
Figure~\ref{all_bv_find_gamma}~(a) and also in Table~\ref{tab_params}.

\begin{figure}
\begin{center}
\includegraphics[width=12cm]{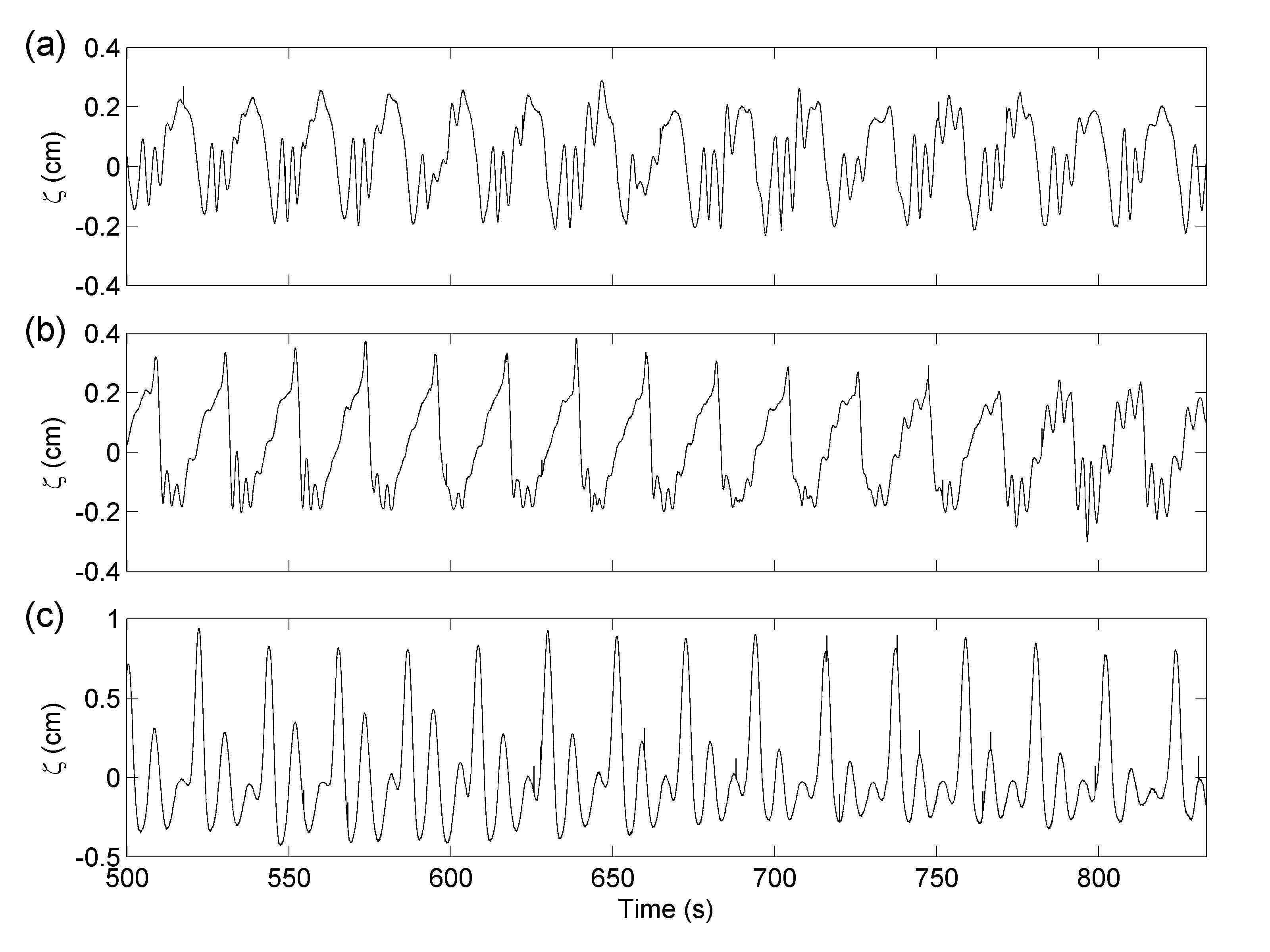}
\end{center}
\caption{Steady-state pycnocline displacements at $x=1.59$~m in (a)
EXP08 ($\gamma=0.11$), (b) EXP22 ($\gamma=0.17$) and (c) EXP16
($\gamma=0.19$).}
\label{pycno_dispts_three_expts}
\end{figure}
\begin{figure}
\begin{center}
\includegraphics[width=12cm]{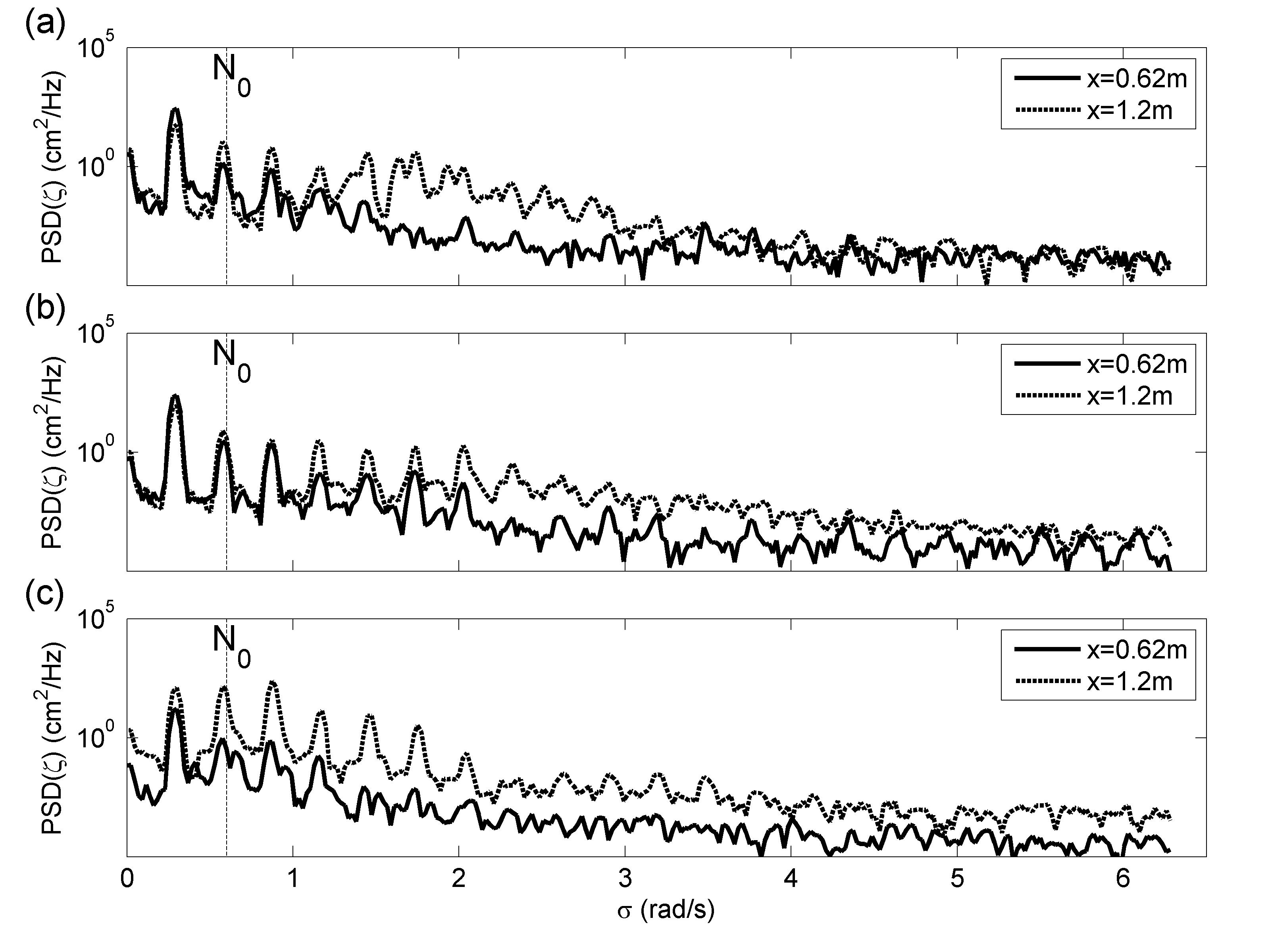}
\end{center}
\caption{Power spectra of the steady-state pycnocline displacements at
 $x=0.62$~m and $x=1.20$~m in (a) EXP08 ($\gamma=0.11$), (b) EXP22 ($\gamma=0.17$) and
(c) EXP16 ($\gamma=0.19$), with all the three plots normalized by the
same constant.}
\label{pycno_spectra_three_expts}
\end{figure}

In figures~\ref{pycno_dispts_three_expts}~(a)-(c), the pycnocline
displacements at $x=1.59$~m from three different experiments (all
corresponding to $\omega_f=0.29$~rad/s) with $\gamma=0.11,0.17,0.19$,
respectively, are presented. As is evident from the figures, groups of
peaked waves are most clearly visible for $\gamma=0.11$ (figure~\ref{pycno_dispts_three_expts}~(a), EXP08). 
The corresponding power spectra, presented in figures~\ref{pycno_spectra_three_expts}~(a)-(c), show that the bulge at
frequencies larger than the forcing frequency is most pronounced for
EXP08, confirming that the bulges in the power spectra are indeed
directly correlated with the appearance of solitons. The bulge is
relatively weaker in EXP22, and is the weakest in EXP16, for which the
higher harmonics dominate over solitary-wave-like features. 
Finally, the solitons being most evident for $\gamma=0.11$ is consistent with the
theoretical criterion of $\gamma\approx0.1$ given by \cite{gerk01}
for optimal excitation of solitons in the pycnocline, and also with oceanic observations.
Indeed, recent studies of the stratification in the Mozambique Channel, where local
generation of ISWs was proposed by~\cite{dasi09}, give $\gamma=0.08$, a value {\color {black} fairly} close
to the observed optimum stratification for soliton generation in our experiments.

The two closely spaced probes at $x=1.2$~m and $x=1.254$~m were
introduced only after EXP08 was performed. Analysing the time series
data from these two closely-spaced acoustic probes in EXP22
($2\omega_f\approx N_0$ and $\gamma=0.17$), we find that the
higher-frequency peaks move with the same speed as the main depression
(or the envelope) on which they sit. They move at $2.3$~cm/s; the 
speed being larger than the mode-1 wave speed ($c_1=1.6$~cm/s, calculated by numerically solving (3.1) with the measured $N(z)$ and $U(z)$ at $x=1.15$~m)for $\omega=2.5$~rad/s 
(frequency representative of the bulge observed in figure \ref{pycno_spectra_three_expts}(b)).
This suggests that the observed solitary waves were mode-1, with the solitons moving 
faster than the corresponding linear mode-1 wave because of nonlinear effects. Finally,
the soliton speed is neither very close to the horizontal phase speed of the
incident wave beam ($1.83$~cm/s, derived from the PIV data, again ruling out
the phase speed matching mechanism in our experiments) nor close to that of the 2nd-mode structure 
of the the 2nd harmonic discussed in section 3 (with observed $\lambda_2=35$~cm, 
we obtain $3.24$~cm/s, suggesting the solitons are rather independent from the higher harmonics).


%% file: 5_discussion.tex
\section{Discussion}


Our experiments were inspired by oceanic observations
(\cite{new92,new02,azev06, dasi07,dasi09}), theoretical works
(\cite{gerk01, akyl07}) and numerical studies (\cite{maug08,gris11}).
Apart from the expected result of the generation of groups of solitary
waves, the full picture that emerges turns out to be intricate, with
the additional features of a mean current and higher harmonics.

\textcolor {black}{We used two measurement techniques, which were complementary. Particle Image Velocimetry measurements were particularly suitable for obtaining data on the harmonics and mean flow in the layer beneath the pycnocline. Ultrasonic probes were used to measure the vertical movement of the pycnocline itself. At the base of the pycnocline there is an overlap: both the techniques provide reliable measurements in this region. A discussion of the consistency between the two techniques is presented in appendix~\ref{appendixA}.}

We have found higher harmonics being generated at the junction of the
incident and reflected beams, as expected from theory by \cite{taba03}. 
Moreover we found that these harmonics are trapped in
the pycnocline when their frequency exceeds, or is close to that of
the constantly stratified lower layer. These trapped harmonics are
each uni-modal, with a well-defined wavelength.  The wavelengths and
vertical profiles of these modes are significantly affected by a
residual current in the upper layer, which, too, finds its origin in
the region where the main beam reflects. Theoretical studies on
internal-wave reflection do predict residual currents~\cite[][for example]{taba05}, 
but only in the very region of reflection. Our
experiments, however, demonstrate that they can be much more extensive
horizontally, in the direction of wave propagation. We are not aware
of any theoretical work that may explain this phenomenon.
{\color {black} The importance of the mean flow on the generation of harmonics, 
and also the apparently three-dimensional structure of the flow, both deserve separate
studies.}

We observe the generation of internal solitary waves only when the
stratification is such that the parameter $\gamma$ is close to $0.1$,
which is consistent with the theoretical criterion put forward by~\cite{gerk01}. Although weak in amplitude, we demonstrated that these
waves are intrinsically nonlinear as they manifest themselves in
spectra not as a superposition of harmonics but as a bulge. This
interpretation of the spectra is suggested by looking at spectra
from a KdV-type model. {\color {black} It is also to be noted that we did not perform experiments at non-zero values smaller than $\gamma=0.11$. We therefore suggest investigations in this regime in future experiments to conclusively establish the optimality of $\gamma=0.1$ for the generation of solitary waves.}

\begin{figure}
\begin{center}
\includegraphics[width=12.5cm]{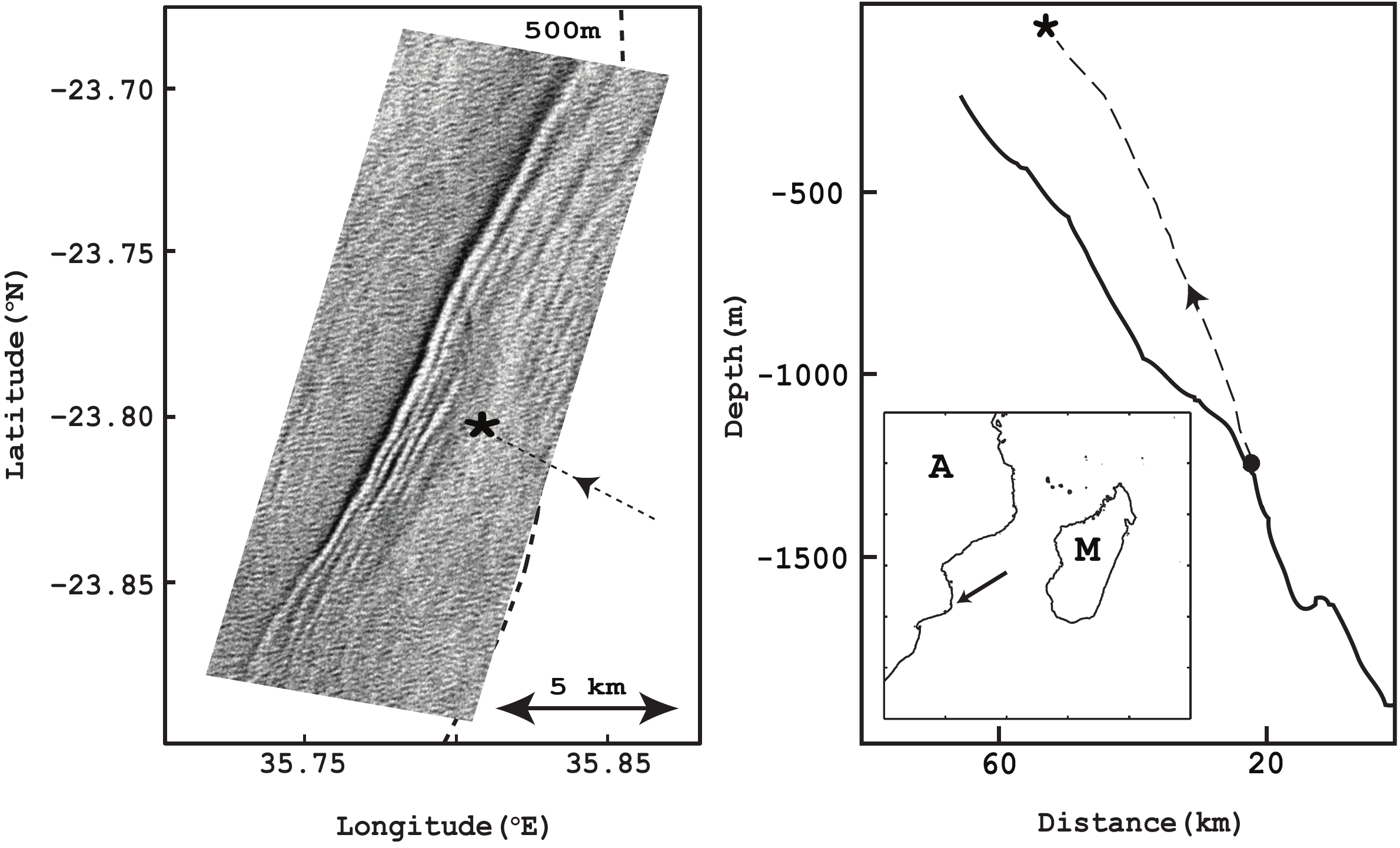}
\end{center}
\caption{(a) SAR image of the Mozambique Channel dated 24 September 2001 at 7:39 UTC,
showing a packet of ISWs consistent with mode-2 vertical structure. The star symbol on the
image indicates the predicted position where an internal tide beam impinges the thermocline.
(b) On the right hand side a vertical section (perpendicular to the ISW crests) is shown with a
simulated beam trajectory based on local stratification (represented by the dashed line). Dark
bold circle shows critical slopes from where the beam admittedly originates. The dark star
symbol shows the location where the beam impacts the thermocline (as well as in part a), and
the arrows indicate the direction of energy propagation. Inset shows geographic location of
the observation; M stands for Madagascar and A for the African continent (the arrow in the
inset denotes the location of the image).}
\label{SAR}
\end{figure}
Although our experiments clearly show the generation of trapped 2nd-mode higher harmonics in the 
pycnocline, the ISWs seem to be of 1st-mode, in line with most oceanic observations. Yet, exceptions may occur,
as is clear from the numerical work by \cite{gris11} and also from oceanic observations such as the one 
shown in Figure 13. Here we see a SAR image from the
Mozambique Channel with surface signatures of ISWs. While these signatures look like many
other typical SAR signatures of ISWs, with larger amplitude waves leading the wave trains in
their direction of propagation, close inspection of their bright and dark patterns reveals that
they {\color {black} are} 
mode-2 solitary like waves, \textcolor {black}{as explained in appendix~\ref{appendixB}}. Indeed, a dark band pattern preceding a bright band
(in the propagation direction) reveals a surface current divergence preceding a convergence,
which is consistent with a mode-2 solitary-like wave \citep{dasi11}. In addition, the ISWs in figure~\ref{SAR} are located close
to (but ahead of) the surfacing of an internal (tidal) ray emanating from critical slopes off the
Mozambique shelf break. The image in Figure~\ref{SAR} provides evidence of mode-2 solitary like
waves that are consistent with Local Generation after the impact of an internal (tide) wave
beam on the pycnocline.

{\color {black} As mentioned in section 1, it was not possible for us to make a direct comparison between our experiments and 
the numerical simulations of \cite{gris11}. The mean flow observed in our experiments, discussed in section 3, 
was not observed in the numerical studies. Since the mean flow affects the phase speeds of the modes, 
it cannot be established from the numerical study how this might affect the criterion for mode-selection proposed by \cite{gris11}. 
The same is true for the higher harmonics. 
So, our laboratory experiments are an inspiration for a next step in the numerical modelling on local generation: 
especially to study the origin and effect of a mean flow.}

Finally, we are not aware of oceanic observations in which trapped higher
harmonics were found in a pycnocline.
In the context of internal tidal beams impinging on a seasonal thermocline, this phenomenon is in any
case not be expected, for the semidiurnal tidal frequency
($1.4\times10^{-4}$ rad/s) is much lower than the typical value of $N$
in the seasonal thermocline ($1\times10^{-2}$ rad/s); in other words,
the harmonic would have to be of an extremely high multiple for it to be trapped.  
The mean flow, on the other hand, has not been observed either, as far as we
are aware, but there seems to be no {\it a priori} reason why it
should not occur.  In any case, in oceanic observations, too, one
would typically find a mixture of higher harmonics (not trapped) and
internal solitary waves.  The idea we have put forward here to
distinguish them in internal-wave spectra, may provide a useful tool
in the oceanographic context as well.

%% file: A_appendix.tex
\section{\textcolor {black}{Comparing PIV and ultrasonic measurements}}
\label{appendixA}


Two different techniques were used to study the internal wave beam impinging on a pycnocline. PIV data from the lower layer of constant stratification provides velocity fields in a vertical plane, with a sampling frequency of $3$~Hz.  An array of ultrasonic probes, on the
other hand, gives times series of local mean vertical displacements of the pycnocline at a much higher sampling frequency of $240$~Hz.
The two techniques are complementary since the PIV measurements offer a large field of view and the ultrasonic probes focus on the pycnocline where PIV cannot be used because of strong optical distortion and sparse particle seeding.
They can also be compared to a certain extent and hence be used to 
reaffirm the conclusions drawn from the data {\color {black} and to demonstrate their consistency}.     

\begin{figure}
\begin{center}
\includegraphics[width=12.5cm]{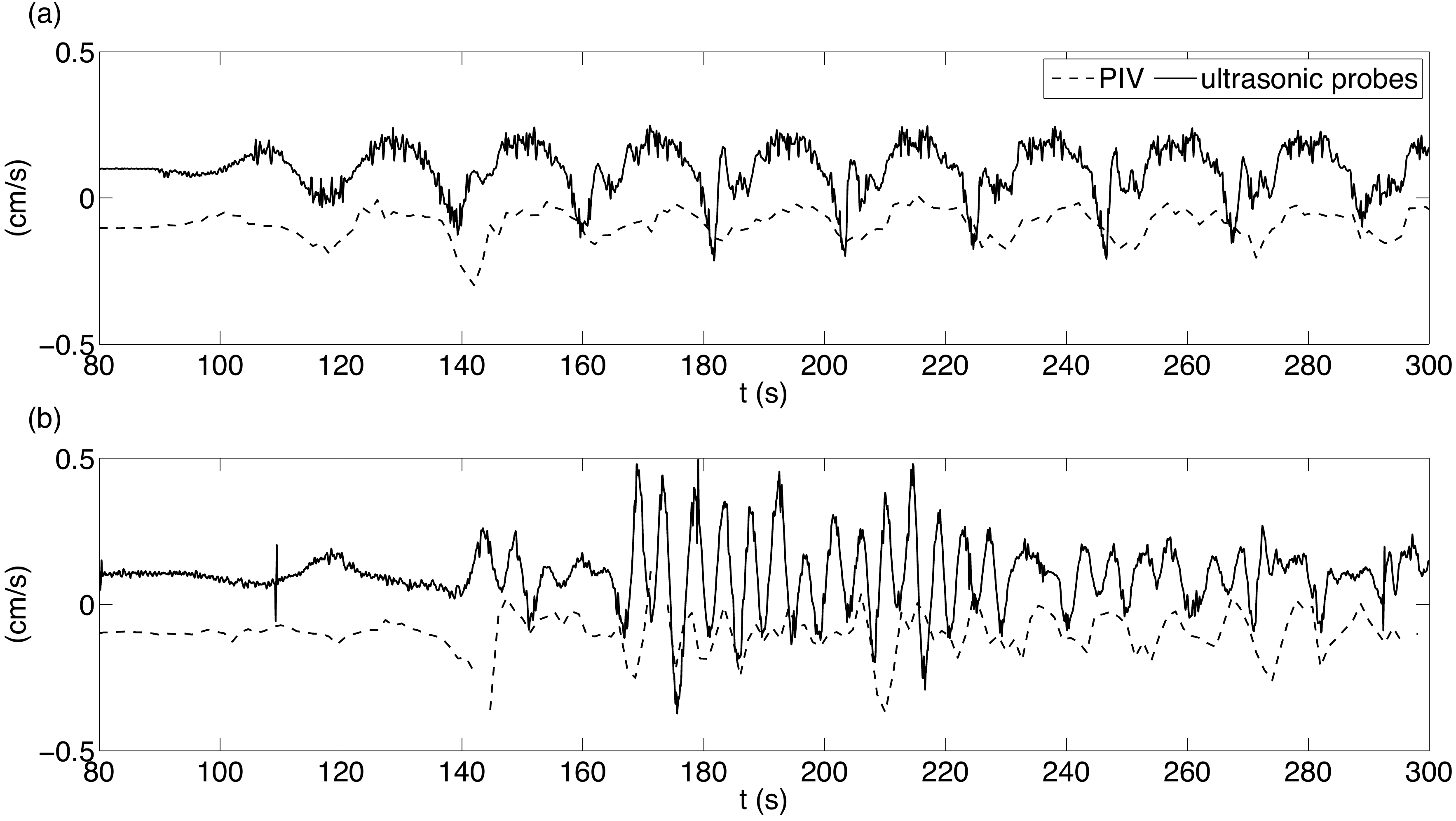}
\end{center}
\caption{EXP 22, comparison of time series of the vertical velocity
$d\zeta/dt$ extracted from the ultrasonic probes and
from the PIV images at the closest location to the ultrasonic probes, for probes located at (a) $x=0.62$~m and (b) $x=1.20$~m. The data do oscillate around $0$ but are arbitrary shifted
of $\pm 0.1$~cm~s$^{-1}$ to make the comparison easier.}
\label{compare_techniques}
\end{figure}

The first time-derivative of the mean pycnocline displacements measured by the ultrasonic probes 
should qualitatively correspond to the vertical velocity obtained from PIV data from just below the
pycnocline. The comparison can only be qualitative because the
acoustic probes provide a vertical average of displacements within the
pycnocline, not a value at a specific depth.
To illustrate this point, we present in Figure\,\ref{compare_techniques} time series of the vertical velocity of
the pycnocline obtained with the two techniques at $x=0.62$~m and
$x=1.2$~m for EXP22. The continuous lines are obtained by taking the
time derivative of the probe signals; the dashed lines represent the
mean vertical velocities (over $5$~mm$^2$) derived from the PIV data centered at these $x$-locations and at a depth $z=-5.25$~cm. 
It is clear that the low-frequency oscillations, of the order of the forcing frequency $\omega_f$, are 
{\color {black} similar} in both datasets, whereas the higher frequencies {\color {black} (i.e.\ larger than $\omega_f$) are more pronounced in}
the ultrasonic probes.
This is mainly due to the fact that these high frequencies are localised in the pycnocline itself
($\omega_f\simeq N_0/2$ for EXP22) and have weak signatures below it.
Finally, an important feature is the presence of out-of-phase oscillations of the vertical velocity of the pycnocline in between its base and its mean position, around $t=170$~s or $220$~s in figure~\ref{compare_techniques}~(b) for instance. These are evidence of mode-2 (or higher) internal waves
in the pycnocline, in agreement with our findings in section 3.2.

%% file: B_appendix.tex
\section{\textcolor {black}{Modal identification from SAR images}}
\label{appendixB}

\begin{figure}
\begin{center}
\includegraphics[width=12.5cm]{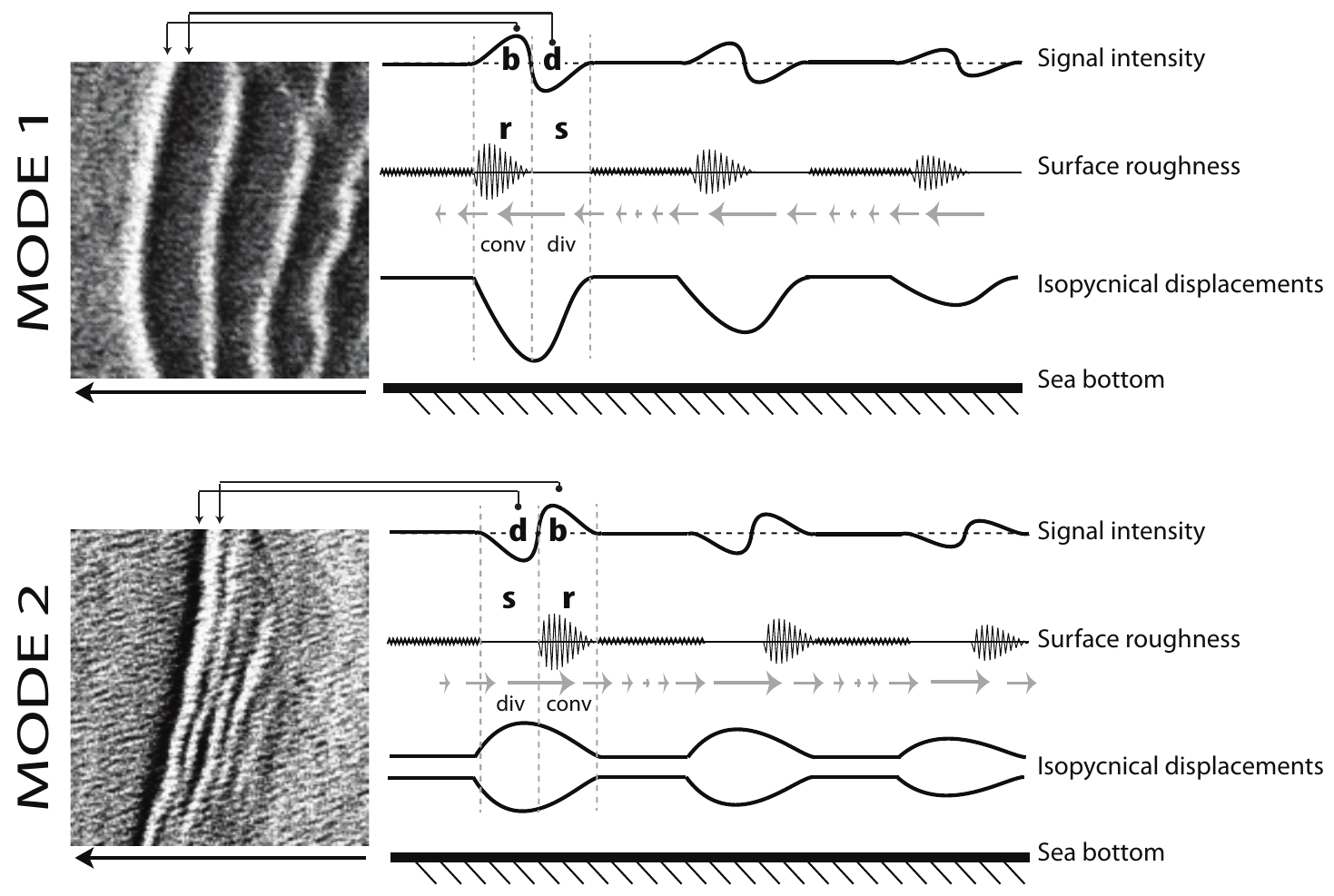}
\end{center}
\caption{Examples of a SAR image with signatures of (a) mode-1 and (b) mode-2 ISWs.}
\label{explain_SAR}
\end{figure}

We present in Figure\,\ref{explain_SAR} two examples of SAR images with signatures of mode-1 and mode-2 ISWs. The internal waves are moving from right to left.

From top to bottom, the horizontal schematic profiles on the right-hand-side of the SAR images represent the following features:
the SAR image intensity profile along ISW propagation direction, with bright enhanced backscatter ({\bf b}) preceding of following dark reduced backscatter ({\bf d}) in the direction of propagation; the surface roughness
representation indicating how rough ({\bf r}) and smooth ({\bf s}) varies along an internal wave train in
relation to isopycnal displacements, below; the surface current variability induced by internal
waves (note the surface field convergence over the leading slopes of ISWs of depression, and
the divergence over the rear slopes); and finally the isopycnal displacements produced by mode-1 or mode-2 internal wave propagation (note the indications of convergence and divergence fields near the surface).

The horizontal profiles represent the same features in both cases. However, one can note how
the SAR signatures, roughness field, current field and convergence and divergence patterns are
reversed for mode-2 ISWs in relation to mode-1 ISWs, leading to a clear discrimination of the modal structure of the propagating ISWs.

%% file: hydra-sol.bbl
\begin{thebibliography}{27}
\expandafter\ifx\csname natexlab\endcsname\relax\def\natexlab#1{#1}\fi

\bibitem[Akylas {\em et~al.\/}(2007)Akylas, Grimshaw, Clarke \& Tabaei]{akyl07}
{\sc Akylas, T.~R., Grimshaw, R. H.~J., Clarke, S.~R. \& Tabaei, A.} 2007
  Reflecting tidal wave beams and local generation of solitary waves in the
  ocean thermocline. {\em J. Fluid Mech.\/} {\bf 593}, 297--313.

\bibitem[Azevedo {\em et~al.\/}(2006)Azevedo, {Da~Silva} \& and]{azev06}
{\sc Azevedo, A., {Da~Silva}, J. C.~B. \& and, A. L.~New} 2006 On the
  generation and propagation of internal solitary waves in the {S}outhern {B}ay
  of {B}iscay. {\em Deep-Sea Res. I\/} {\bf 53}, 927--941.

\bibitem[{Da~Silva} {\em et~al.\/}(2007){Da~Silva}, New \& Azevedo]{dasi07}
{\sc {Da~Silva}, J. C.~B., New, A.~L. \& Azevedo, A.} 2007 On the role of {SAR}
  for observing {"local generation"} of internal solitary waves off the
  {I}berian {P}eninsula. {\em Can. J. Remote Sensing\/} {\bf 33}~(5), 388--403.

\bibitem[{Da~Silva} {\em et~al.\/}(2009){Da~Silva}, New \& Magalhaes]{dasi09}
{\sc {Da~Silva}, J. C.~B., New, A.~L. \& Magalhaes, J.~M.} 2009 Internal
  solitary waves in the {M}ozambique {C}hannel: observations and
  interpretation. {\em J. Geophys. Res.\/} {\bf 114}~(C05001),
  doi:10.1029/2008JC005125.

\bibitem[{Da~Silva} {\em et~al.\/}(2011){Da~Silva}, New \& Magalhaes]{dasi11}
{\sc {Da~Silva}, J. C.~B., New, A.~L. \& Magalhaes, J.~M.} 2011 On the
  structure and propagation of internal solitary waves generated at the
  {M}ascarene {P}lateau in the {I}ndian {O}cean. {\em Deep-Sea Res. I\/} {\bf
  58}~(3), 229--240.

\bibitem[Delisi \& Orlanski(1975)]{deli75}
{\sc Delisi, D.~P. \& Orlanski, I.} 1975 On the role of density jumps in the
  reflexion and breaking of internal gravity beams. {\em J. Fluid Mech.\/} {\bf
  69}, 445--464.

\bibitem[Fincham \& Delerce(2000)]{finc00}
{\sc Fincham, A. \& Delerce, G.} 2000 Advanced optimization of correlation
  imaging velocimetry algorithms. {\em Experiments in Fluids\/} {\bf 29}, S1.

\bibitem[Gerkema(1996)]{gerk96}
{\sc Gerkema, T.} 1996 A unified model for the generation and fission of
  internal tides in a rotating ocean. {\em J. Mar. Res.\/} {\bf 54}~(3),
  421--450.

\bibitem[Gerkema(2001)]{gerk01}
{\sc Gerkema, T.} 2001 Internal and interfacial tides: beam scattering and
  local generation of solitary waves. {\em J. Mar. Res.\/} {\bf 59}~(2),
  227--255.

\bibitem[Gostiaux \& Dauxois(2007)]{gost07b}
{\sc Gostiaux, L. \& Dauxois, T.} 2007 Laboratory experiments on the generation
  of internal tidal beams over steep slopes. {\em Phys. Fluids\/} {\bf 19},
  028102.

\bibitem[Gostiaux {\em et~al.\/}(2007)Gostiaux, Didelle, Mercier \&
  Dauxois]{gost07a}
{\sc Gostiaux, L., Didelle, H., Mercier, S. \& Dauxois, T.} 2007 A novel
  internal waves generator. {\em Exp. in Fluids\/} {\bf 42}, 123--130.

\bibitem[Grisouard {\em et~al.\/}(2011)Grisouard, Staquet \& Gerkema]{gris11}
{\sc Grisouard, N., Staquet, C. \& Gerkema, T.} 2011 Generation of internal
  solitary waves in a pycnocline by an internal wave beam: a numerical study.
  {\em J. Fluid Mech.\/} {\bf 676}, 491--513.

\bibitem[Jackson(2007)]{jack07}
{\sc Jackson, C.} 2007 Internal wave detection using the moderate resolution
  imaging spectroradiometer (modis). {\em J. Geophys. Res.\/} {\bf 112},
  C11012.

\bibitem[Jiang \& Marcus(2009)]{jian09}
{\sc Jiang, C.~H. \& Marcus, P.~S.} 2009 Selection rules for the nonlinear
  interaction of internal gravity waves. {\em Phys. Rev. Letters\/} {\bf 102},
  124502.

\bibitem[King {\em et~al.\/}(2009)King, Zhang \& Swinney]{king09}
{\sc King, B., Zhang, H.~P. \& Swinney, H.~L.} 2009 Tidal flow over
  three-dimensional topography in a stratified fluid. {\em Phys. Fluids\/} {\bf
  21}, 116601.

\bibitem[LeBlond \& Mysak(1978)]{lebl78}
{\sc LeBlond, P.~H. \& Mysak, L.~A.} 1978 {\em Waves in the ocean\/}. Elsevier,
  Amsterdam.

\bibitem[Mathur \& Peacock(2009)]{math09}
{\sc Mathur, M. \& Peacock, T.} 2009 Internal wave beam propagation in
  nonuniform stratifications. {\em J. Fluid Mech.\/} {\bf 639}, 133--152.

\bibitem[Maug\'e \& Gerkema(2008)]{maug08}
{\sc Maug\'e, R. \& Gerkema, T.} 2008 Generation of weakly nonlinear
  nonhydrostatic internal tides over large topography: a multi-modal approach.
  {\em Nonlin. Process. Geophys.\/} {\bf 15}, 233--244.

\bibitem[Mercier {\em et~al.\/}(2010)Mercier, Martinand, Mathur, Gostiaux,
  Peacock \& Dauxois]{merc10}
{\sc Mercier, M.~J., Martinand, D., Mathur, M., Gostiaux, L., Peacock, T. \&
  Dauxois, T.} 2010 New wave generation. {\em J. Fluid Mech.\/} {\bf 657},
  308--334.

\bibitem[Michallet \& Barth\'{e}lemy(1997)]{mich97}
{\sc Michallet, H. \& Barth\'{e}lemy, E.} 1997 Ultrasonic probes and data
  processing to study interfacial solitary waves. {\em Exp. in Fluids\/} {\bf
  22}, 380--386.

\bibitem[New \& {Da~Silva}(2002)]{new02}
{\sc New, A.~L. \& {Da~Silva}, J. C.~B.} 2002 Remote-sensing evidence for the
  local generation of internal soliton packets in the central {B}ay of
  {B}iscay. {\em Deep-Sea Res.\/} {\bf 49}~(5), 915--934.

\bibitem[New \& Pingree(1990)]{new90}
{\sc New, A.~L. \& Pingree, R.~D.} 1990 Large-amplitude internal soliton
  packets in the central {B}ay of {B}iscay. {\em Deep-Sea Res.\/} {\bf 37},
  513--524.

\bibitem[New \& Pingree(1992)]{new92}
{\sc New, A.~L. \& Pingree, R.~D.} 1992 Local generation of internal soliton
  packets in the central {B}ay of {B}iscay. {\em Deep-Sea Res.\/} {\bf 39},
  1521--1534.

\bibitem[Tabaei \& Akylas(2003)]{taba03}
{\sc Tabaei, A. \& Akylas, T.~R.} 2003 Nonlinear internal gravity wave beams.
  {\em J. Fluid Mech.\/} {\bf 482}, 141--161.

\bibitem[Tabaei {\em et~al.\/}(2005)Tabaei, Akylas \& Lamb]{taba05}
{\sc Tabaei, A., Akylas, T.~R. \& Lamb, K.~G.} 2005 Nonlinear effects in
  reflecting and colliding internal wave beams. {\em J. Fluid Mech.\/} {\bf
  526}, 217--243.

\bibitem[Thomas \& Stevenson(1972)]{thom72}
{\sc Thomas, N.~H. \& Stevenson, T.~N.} 1972 A similarity solution for viscous
  internal waves. {\em Journal of Fluid Mechanics\/} {\bf 54}, 495--506.

\bibitem[Thorpe(1998)]{thor98}
{\sc Thorpe, S.~A.} 1998 Nonlinear reflection of internal waves at a density
  discontinuity at the base of the mixed layer. {\em J. Phys. Oceanogr.\/} {\bf
  28}, 1853--1860.

\end{thebibliography}
